\documentclass[pra,showpacs]{revtex4}
\usepackage{graphicx}
\usepackage{bm}

\begin{document}

\def\bra{\langle}  \def\ket{\rangle}
\def\ketpsi{| \psi \rangle}


\title{Appearance and Stability of Anomalously Fluctuating States\\ 
in Shor's Factoring Algorithm
}

\author{Akihisa Ukena}
\email{ukena@ASone.c.u-tokyo.ac.jp} 

\author{Akira Shimizu}
\email{shmz@ASone.c.u-tokyo.ac.jp}
\altaffiliation{also at 
PRESTO, JST, 4-1-8 Honcho Kawaguchi, Saitama, Japan
}

\affiliation{
Department of Basic Science, University of Tokyo, 
3-8-1 Komaba, Tokyo 153-8902, Japan}

\date{1 August 2003}

\begin{abstract}
We analyze quantum computers which perform Shor's factoring 
algorithm, paying attention to asymptotic properties as 
the number $L$ of qubits is increased.
Using numerical simulations and a general theory of the stabilities of 
many-body quantum states, we show the following:
Anomalously fluctuating states (AFSs), which have anomalously large 
fluctuations of additive operators, appear in various stages of 
the computation. For large $L$, they decohere at anomalously 
great rates by weak noises that simulate noises in real systems.
Decoherence of some of the AFSs is fatal to the results of the computation, 
whereas decoherence of some of the other AFSs does not have
strong influence on the results of the computation.
When such a crucial AFS decoheres, 
the probability of getting the correct computational result is 
reduced approximately proportional to $L^2$.
The reduction thus becomes anomalously large
with increasing $L$, even when the coupling constant to the noise
is rather small. 
Therefore, quantum computations 
should be improved in such a way that
all AFSs appearing in the algorithms do not 
decohere at such great rates in the existing noises.
\end{abstract}

\maketitle

\section{Introduction}\label{sec-intro}

Quantum computers are considered to be more efficient 
than classical computers in solving certain problems 
\cite{Deutsch_1,Shor_1,Ekert_Jozsa,Nielsen_Chaung}.
Since the efficiency of computation becomes relevant only when 
the size $N$ of the input is huge, 
the efficiency is defined in terms 
of the asymptotic behavior of the number $Q$ of the 
computational steps as the size $N$ of the input
is increased.
To study the asymptotic behavior, one must take the data size 
large but finite.
Since larger $N$ generally requires a larger 
number $L$ of qubits as $L \sim \log N$, 
$L$ of a relevant quantum computer becomes large but finite.
Therefore, a relevant quantum computer is a 
many-body quantum system with large but finite 
degrees of freedom, $1 \ll L < +\infty$.

In the conventional many-body physics,  
one is usually interested in states 
that approach 
as $L \to \infty$
 a ``vacuum'' state and finite excitations on it.
This means that one is usually uninterested in 
other states of finite systems.
In a quantum computer, on the other hand,
various states are generated
according to the algorithm and the input.
It is therefore expected that some of them would be very different from 
states that are treated in the conventional many-body physics.
It is very interesting to reveal physical properties 
of such ``anomalous'' states as well as their roles in quantum computations.

For quantum states of 
general systems with large but finite degrees of freedom, 
Shimizu and Miyadera (SM) recently 
studied the stabilities
against weak noises, against weak perturbations from environments, 
and against local measurements \cite{SM02}.
By fully utilizing the locality of the theory, 
they obtained the general and universal results:
the stabilities 
of quantum states are determined by long-distance 
correlations between local operators. 
As measures of the long-range correlations of quantum states, 
SM employed the ``cluster property,''
which plays a fundamental role in field theory \cite{haag}, 
and the ``fluctuations of 
additive operators,'' 
which will be explained in the following section.
If a pure state 
has anomalously large fluctuation of an additive operator(s),
then the state 
has a long-distance correlation(s).
Such a state does not have the cluster property, 
and is quite anomalous in many-body physics.
Such anomalies are directly related to the stabilities 
of the quantum states.

Since the stability of quantum states  
against noises has been considered as a key to realizing
quantum computers \cite{unruh,Palma,Miquel_Paz_Perazzo}, 
it is interesting to apply the general theory by SM to quantum computers.
In this paper, we analyze quantum computers performing Shor's factoring 
algorithm \cite{Shor_1,Ekert_Jozsa,Nielsen_Chaung}
using the general theory by SM and numerical simulations.
We show that anomalous states, which have anomalously large 
fluctuations of additive operators, appear during the computation.
For large $L$, they decohere at anomalously great rates in the presence of 
long-wavelength noises, 
while for small $L$ the decoherence rates are of the same order of 
magnitude as those of normal states.
The decoherence of some of the anomalous states, not all of them, 
results in the reduction of the success probability of the computation.
Therefore, the decoherence of such anomalous states is 
crucial to Shor's factoring algorithm with huge inputs.

\section{Summary of the general theory of the 
decoherence rates of
quantum states in systems of large but finite degrees of freedom}
\label{sec-SM}

SM \cite{SM02} studied stabilities of quantum states of
general systems of large but finite degrees of freedom.
Among three kinds of stabilities discussed by them, 
we focus on the stability against weak classical noises; namely, we focus on the decoherence due to weak classical noises.

As compared with general many-body systems, 
quantum computers are often assumed to have 
the following special properties
(although they are not necessary):
(a) The system Hamiltonian is assumed to be negligible, 
so that quantum states of a quantum computer do not 
evolve unless the computer is subjected to external operations and/or noises.
(b) Qubits are assumed to be located on a one-dimensional lattice. 
In this section, 
we summarize the result of SM
for the decoherence rate $\Gamma$, 
assuming these special properties.

\subsection{Normalized additive operator}

We consider 
a quantum computer that is composed of $L$ ($\gg 1$) qubits
which are located on sites of a one-dimensional lattice with a unit 
lattice constant.
Let $\hat a(\ell)$ be a {\em local operator} at site $\ell$
($= 1, 2, \cdots, L$), 
which for qubit systems is a polynomial of the Pauli operators
$\hat \sigma_x(\ell), \hat \sigma_y(\ell), \hat \sigma_z(\ell)$, 
acting on the qubit at $\ell$.
Such a polynomial becomes a linear combination of 
the identity operator $\hat 1(\ell)$ and 
$\hat \sigma_x(\ell), \hat \sigma_y(\ell), \hat \sigma_z(\ell)$
because of the SU(2) algebra.
We define a {\em normalized additive operator} $\hat A$ by
\begin{equation}
\hat A
=
\frac{1}{L}\sum_{\ell = 1}^L \hat a(\ell),
\label{A}\end{equation}
which is the normalized one of the ``additive operator'' defined 
in Ref.\ \cite{SM02}.
For example, if we take
\begin{equation}
\hat a(\ell) = (-1)^{\ell} \hat \sigma_z(\ell),
\label{smsigma}\end{equation}
then $\hat A$ is the $z$ component of the ``staggered magnetization,'' 
which has a finite expectation value 
when the system has an antiferromagnetic order.

Note that 
normalized additive operators are
macroscopic operators.
Thermodynamics assumes that fluctuations of 
any macroscopic observables are $o(V^2)$ for pure phases \cite{SM02}, 
where $V$ is the volume of the system.
However, this is {\em not} necessarily satisfied 
by pure quantum states of finite macroscopic systems 
\cite{SM02,HL,KT,pre01}, 
as will be described in the following.

\subsection{$L$ dependence of quantum states}

In order to discuss $L$ dependences of properties of quantum states, 
some rule is necessary that defines $L$ dependence of quantum states.
The simplest one of such a rule is 
that the quantum states are homogeneously extended with increasing $L$.
For example, consider 
a superposition of two N\'{e}el states,
\begin{equation}
\frac{1}{\sqrt{2}}
|1010 \cdots 10 \rangle
+
\frac{1}{\sqrt{2}}
|0101 \cdots 01 \rangle,
\label{catNeel}\end{equation}
where $|1\rangle$ and $|0\rangle$ denote the spin-up and -down states, 
respectively.
To increase $L$ of this state, one can simply add two spins as
$ 
\frac{1}{\sqrt{2}}
|1010 \cdots 1010 \rangle
+
\frac{1}{\sqrt{2}}
|0101 \cdots 0101 \rangle.
$ 
Unfortunately, 
states in quantum computers do not have such simple homogeneity.
However, they are homogeneous in a broad sense because
states with a larger $L$ (which is necessary for a larger input $N$) 
and states with a smaller $L$
are both generated according to the same algorithm.
As a result, both of them have similar structures, 
as will be demonstrated explicitly in Sec. \ref{ss-similarity}.
This allows us to 
analyze the asymptotic behaviors of properties of 
states in quantum computers as $N \to \infty$.

\subsection{Normally and anomalously fluctuating states}
\label{ss-NFSAFS}

As a measure of correlations between distant qubits for a system of 
many qubits, 
SM \cite{SM02} proposed the use of 
fluctuations of additive operators.
We here summarize their proposal 
in terms of {\em normalized} additive operators.

Consider a pure state $\ketpsi$, and put
\begin{equation}
\Delta \hat A
\equiv
\hat A - \langle \psi | \hat A | \psi \rangle.
\end{equation}
We focus on the $L$ dependence of the fluctuation 
$\langle \psi | (\Delta \hat A)^2 | \psi \rangle$
for $L \gg 1$, and define an index $p$ by
\begin{equation}
\langle \psi | (\Delta \hat A)^2 | \psi \rangle
=O(L^{p-2}).
\label{def-p}\end{equation}
The value of $p$ depends on both $\hat A$ and $| \psi \rangle$.
For example, 
if the correlation between $\hat a(\ell)$ and $\hat a(\ell')$ 
for large $|\ell-\ell'|$ 
is negligibly small for $| \psi \rangle$, 
then $p=1$
for the normalized additive operator 
$\hat A$ which is composed of $\hat a(\ell)$ as Eq.\ (\ref{A}). 

For a given $| \psi \rangle$, 
if the maximum value of $p$ (among those of
all normalized additive operators)
is unity, 
$| \psi \rangle$ is called a normally fluctuating state (NFS).
It is easy to show that any separable state is a NFS.
Note however that 
{\em the inverse is not necessarily true}.
For example, 
$ 
|100 \cdots 0 \ket,\
|010 \cdots 0 \ket,\
\cdots,\
|000 \cdots 1 \ket
$ 
are all separable states, hence are NFSs.
Their superposition
\begin{equation}
\frac{1}{\sqrt{L}}
\left[
|100 \cdots 0 \ket
+|010 \cdots 0 \ket
+|001 \cdots 0 \ket
+ \cdots
+|000 \cdots 1 \ket
\right]
\equiv | {\rm W} \rangle
\label{W}\end{equation}
is also a NFS because $p=1+o(L)/L$, but 
non separable.

On the other hand, 
if there is a normalized additive operator(s) $\hat A$
for which $p=2$, 
then the pure state 
$| \psi \rangle$ is called an anomalously fluctuating state (AFS)
because the fluctuation of $\hat A$ is anomalously large.
In this case, 
$\hat a(\ell)$ and $\hat a(\ell')$, 
which compose $\hat A$,  
are strongly correlated even when $|\ell-\ell'| \sim L$.
Since AFSs are pure states, 
this indicates that 
AFSs are entangled {\em macroscopically} \cite{SM02}.
This entanglement is macroscopic because
one cannot turn a NFS into an AFS by 
adding a small ($\sim L^0$) number of Bell pairs.
Note, in particular, that 
the state $| {\rm W} \rangle$ of Eq.~(\ref{W}) 
{\em is} entangled, but {\em not} macroscopically entangled, 
hence is not an AFS but a NFS.
In this way, the value of $p$ 
can be taken as
a quantitative measure of {\em macroscopic} entanglement.

A simple example of an AFS is
the state of Eq.\ (\ref{catNeel}), 
for which 
$ 
\langle \psi | (\Delta \hat A)^2 | \psi \rangle
= 1
$ 
for the staggered magnetization defined by Eqs.\ (\ref{A}) and 
(\ref{smsigma}).
Another simple example of an AFS is
\begin{equation}
\frac{1}{\sqrt{2}}
|000 \cdots 0 \rangle
+
\frac{1}{\sqrt{2}}
|111 \cdots 1 \rangle
\equiv | {\rm C} \rangle,
\label{catFerro}\end{equation}
for which the fluctuation of the ``magnetization'' 
$ 
\hat M_z
\equiv
\frac{1}{L} \sum_\ell \hat \sigma_z(\ell)
$ 
is anomalously large;
$ 
\langle {\rm C} | (\Delta \hat M_z)^2 | {\rm C} \rangle
= 1.
$ 

From the viewpoints of many-body physics
and experiments, 
the index $p$ seems a natural measure of macroscopic entanglement.
In ferromagnets, for example, $| {\rm C} \rangle$ is 
quite anomalous because it is a superposition of two states
that have different values of the macroscopic variable $\hat M_z$.
Such a state is usually discarded in many-body physics for many 
reasons. One reason is that 
it is very hard to generate such a state experimentally.
Another reason is that such a state is not allowed as a
pure state in the limit of infinite degrees of freedom, 
$L \to \infty$ \cite{SM02}.
On the other hand, 
$| {\rm W} \rangle$ is a normal state which can easily be 
generated experimentally, 
although it is sometimes 
classified as a strongly entangled state in quantum information theory.
In insulating solids, for example, 
the state vector of a many-body state in which 
a Frenkel exciton is excited on the ground state 
takes the form of $| {\rm W} \rangle$; namely, 
in many-body physics $| {\rm W} \rangle$ is just an ordinary state 
in which a single quasiparticle is excited on the ground state.
Since many-body physics is directly related to 
(and has been tested by)
experiments, a normal (abnormal) state in many-body physics is also 
a normal (abnormal) state in experiments.
Therefore, our measure of macroscopic entanglement seems natural
from the viewpoints of many-body physics and experiments.
Considering that quantum computers should be fabricated from 
real materials, this indicates also that our measure would be 
suitable for discussing realization of quantum computers.
Furthermore, 
an efficient method of computing $p$ for general states 
has been developed in Ref.~\cite{SS03}, 
whereas some of the other measures are hard to compute for general states.
This is also an advantage of the present measure  \cite{disadv}.

Although states with $1<p<2$ are possible \cite{NGmodes}, 
we focus on two classes of states, 
NFSs ($p=1$) and AFSs ($p=2$), in this paper.

\subsection{Decoherence rate}
\label{ss-dr}

We consider the decoherence rate $\Gamma$ of states of quantum computers.
For the origin of the decoherence, 
we assume a weak classical noise
$f(\ell, t)$, acting on every site $\ell$,
with vanishing average $\overline{f(\ell,t)} = 0$.
The case of a weak perturbation from the environment
can be treated in a similar manner. 
We assume that 
$\overline{f(\ell,t) f(\ell',t')}$ 
[$=\overline{f(\ell',t') f(\ell,t)}$] 
depends only on
$|\ell-\ell'|$ and $|t-t'|$.
We denote the spectral intensity of $f(\ell,t)$ by $g(k,\omega)$,
which is non-negative by definition.
Here, $k$ takes discrete values from $-\pi$ to $\pi$ 
with separation $2 \pi/L$ \cite{k},
whereas $\omega$ takes continuous values.
The autocorrelation function can be expressed as
\begin{equation}
\overline{f(\ell,t) f(\ell',t')}
=
\sum_k \int \frac{d \omega}{2 \pi} g(k,\omega)
e^{i k (\ell-\ell') - i \omega (t-t')}.
\end{equation}

Since physical interactions must be local, 
the interaction between the qubit system and the noise 
should be the sum of local interactions:
\begin{equation}
\hat H_{\rm int}(t)
=
\lambda
\sum_{\ell=1}^L
f(\ell,t) \hat a(\ell),
\label{int-noise}\end{equation}
where $\lambda$ is a positive constant and $\hat a(\ell)$ is a local operator at site $\ell$.
Since increasing $\lambda$ is equivalent to 
increasing the amplitude of the noise, 
$\lambda$ may be interpreted as the 
noise amplitude times the coupling constant.
We shall therefore 
take $f$ to have a normalized amplitude, as Eq.\ (\ref{f(t)}), 
and vary the noise amplitude by varying $\lambda$.

Assuming this local interaction and 
a short correlation time for the noise, 
SM showed that for $L \gg 1$
the decoherence rate $\Gamma$ of a pure state $\ketpsi$ 
is directly related to 
fluctuations of additive operators.
In terms of {\em normalized} additive operators, 
their formula reads
\begin{equation}
\Gamma
\simeq
\lambda^2 L^2 \sum_k
g(k)
\langle \psi | \Delta \hat A_k^\dagger \Delta \hat A_k | \psi \rangle.
\label{r-noise-2}\end{equation}
Here, 
$g(k)$ is an average value of $g(k,\omega)$ \cite{SM02} 
and
$
\Delta \hat A_k
\equiv
\hat A_k - \langle \psi | \hat A_k | \psi \rangle
$,
where
\begin{equation}
\hat A_k
\equiv
\frac{1}{L}\sum_{\ell=1}^L \hat a(\ell) e^{-i k \ell}.
\label{Ak}\end{equation}
For quantum computers, $g(k) \simeq g(k,0)$  because 
the system Hamiltonian is negligible.
Note that $\hat A_k$ is a normalized additive operators, 
where $\hat a(\ell) e^{-i k \ell}$ of Eq.\ (\ref{Ak}) corresponds to 
$\hat a(\ell)$ of Eq.\ (\ref{A}).
Hence, formula (\ref{r-noise-2}) shows that 
$\Gamma$ of a pure state 
$\ketpsi$ of a quantum computer is determined by 
the fluctuation of a normalized additive operator
that is composed of the local operators in $\hat H_{\rm int}$.
Note that the formula of a pioneering work by Palma {\it et al}.\ \cite{Palma} 
is a special case of the above general formula.

\subsection{Fragility}

We say that a quantum state is ``fragile'' if its decoherence rate $\Gamma$ is 
anomalously great in such a way that 
\begin{equation}
\Gamma \sim 
K L^{1+\delta},
\label{KN1pd}\end{equation}
where $K$ is a function of microscopic parameters,
such as $\lambda$,
and $\delta$ is a positive constant.
This is an anomalous situation in which 
the decoherence rate {\em per qubit}, 
$
\Gamma/L \sim K L^{\delta}
$,
grows with increasing $L$.
For a nonfragile state (for which $\delta=0$), in contrast, 
$
\Gamma/L \sim K
$
is independent of $L$, 
in consistency with the naive expectation.
This is a normal situation in which 
the total decoherence rate $\Gamma$ is simply the sum of 
local decoherence rates.

For large $L$, 
fragile quantum states 
decohere much faster than 
nonfragile states
because $\Gamma$ becomes anomalously great, 
even when the coupling constant between 
the system and the noise is small.

\subsection{Non fragility of all NFSs and fragility of some AFSs}
\label{ss-frg-of-AFS}

When $| \psi \rangle$ is a NFS, 
$\langle \psi | \Delta \hat A_k^\dagger \Delta \hat A_k | \psi \rangle
\leq O(1/L)$
for any $\hat A_k$, hence
\begin{equation}
\Gamma
\lesssim \lambda^2
O(L) \sum_k g(k).
\label{r-noise-NFS}\end{equation}
Since 
$
\sum_k g(k)
= \int \overline{f(x,t)f(x,0)} {\rm d} t
$
does not depend on $L$, 
we find that 
NFSs never become fragile under  
weak perturbations from any random noises \cite{SM02}.

When $| \psi \rangle$ is an AFS, 
on the other hand, 
$\langle \psi | \Delta \hat A_k^\dagger \Delta \hat A_k | \psi \rangle
= O(L^0)$
for some $\hat A_k$, i.e., for some local operator 
$\hat a(x)$ and some wave number $k=k_0$.
Hence, if $\hat H_{\rm int}$ has a term that 
is composed of such $\hat a(x)$'s, then
\begin{equation}
\Gamma
\simeq
\lambda^2 O(L^2) g(k_0)+
\lambda^2 O(L) \sum_{k (\neq k_0)} g(k).
\label{r-noise-AFS}\end{equation}
In this case, the AFS becomes fragile if 
$g(k_0) = O(L^{-1+\delta})$, where $\delta >0$; namely, for an AFS, there {\em can} exist 
weak classical noises or weak perturbations from environments
that make the AFS fragile.
Whether such a noise or environment {\em really} exists
in a quantum computer system depends on physical situations. 

Although SM also considered a more fundamental instability
of AFSs (i.e., instability against local measurements), 
the fragility is sufficient for the purpose of the present paper.

\section{Conjectures on quantum computers}\label{sec-conj}

From the general results summarized in the preceding section, 
we are led to the following conjectures on quantum computers\cite{S_talk}.

\subsection{Quantum computers should utilize AFSs}

Entanglement has been considered to be essential to 
efficient quantum computations 
\cite{Deutsch_1,Shor_1,Ekert_Jozsa,Nielsen_Chaung,miyake,Jozsa_Linden}.
To discuss the efficiency of computation,
one must study the asymptotic behavior of the 
computational time as $N \to \infty$.
It is natural to consider that more entanglement is 
required for larger $N$.
Since AFSs are entangled macroscopically, 
we are led to the following conjecture, 
which we call conjecture (i): 
In performing an algorithm that is much more efficient than 
any classical algorithms, 
a quantum computer should utilize AFSs 
in some stages of the computation.

Although the use of entanglement in quantum computations was 
somehow confirmed 
in the previous works \cite{miyake,Jozsa_Linden}, 
the magnitude of entanglement for large $L$ can be defined in various ways.
As a quantitative measure of entanglement, 
we propose here to use the asymptotic behavior,
Eq.\ (\ref{def-p}), 
of fluctuations of normalized additive operators.
This enables us not only to identify macroscopically entangled states
for large $L$, but also to estimate their
decoherence rates using formulas (\ref{r-noise-2}), 
(\ref{r-noise-NFS}), and (\ref{r-noise-AFS}).

\subsection{Some of AFSs appearing during quantum computation would be fragile}
\label{ss-emn}

If conjecture (i) is the case, 
the results of Sec. \ref{ss-frg-of-AFS}
lead us to the second conjecture,
which we call conjecture (ii): 
Some of the AFSs appearing during quantum computation would be fragile
under realistic weak classical noises or 
weak perturbations from environments.

For example, suppose that electromagnetic noises at 4 K is 
the dominant noise.
If the physical dimension of a quantum computer is less than 1 cm, 
then $g(k)$ of the electromagnetic noises behaves as follows \cite{k}:
\begin{equation}
g(k) \simeq 
\left\{
\begin{array}{ll}
O(L^0) & (k=0)
\\
O(1/L) & (k \gg 2 \pi/L).
\end{array}
\right.
\label{longwlnoise}\end{equation}
Therefore, 
according to Eq.\ (\ref{r-noise-AFS}), 
AFSs with $k_0 = 0$ becomes fragile (with $\delta =1$), 
whereas 
AFSs with $k_0 \gg 2 \pi/L$ are nonfragile; namely, if AFSs with $k_0 = 0$ appear during the computation, 
they are fragile in electromagnetic noises at 4 K.

\subsection{Fragility of some AFSs should be fatal to quantum computation} 

If a fragile AFS is used in the quantum computation, 
it decoheres at an anomalously great rate as given by 
Eq.\ (\ref{KN1pd}) with $\delta > 0$.
Since quantum coherence is considered to be important in 
quantum computations,
we are further led to the third conjecture, which we call
conjecture (iii):
The anomalously fast decoherence of such a fragile AFS(s) 
should be fatal to a quantum computation, 
and the quantum computation would become impossible 
for large $L$ even if 
the coupling constant between 
the system and the noise or environment is small, 
unless 
error correcting codes (ECC) \cite{ec1,ec2,ec3,ec4} and/or
a decoherence-free subspace (DFS) \cite{dfs1,dfs2,dfs3}
were successfully used.
Since such efficient ECC and/or a DFS
against all existing noises
would not be easy to realize (see Sec. \ref{ss-ECandDFS}), 
improvements of the algorithm seem necessary; namely, if possible, 
one should implement the algorithm in such a way that 
only AFSs that are not fragile in existing noises are used. 

In this paper, we study these conjectures by analyzing 
quantum computers performing Shor's factoring algorithm.

\section{Quick summary of Shor's factoring algorithm}
\label{sec-Shor}

To study the conjectures raised in the preceding section, 
we perform numerical simulations on Shor's factoring algorithm
\cite{Shor_1,Ekert_Jozsa}, 
which is believed to be much (exponentially) faster than any
classical algorithms.
In order to establish notations, 
we summarize in this section
the main part,
which we will simulate, of the algorithm.

In order to factor an integer $N$, 
we use two quantum registers $R_{1}$ and $R_{2}$, 
which are called the first and second registers, respectively.
They are composed of $L_1$ and $L_2$ qubits, respectively, where
\begin{eqnarray}
&& 2 \log N \leq L_1 < 2 \log N + 1,
\\
&& \log N \leq L_2 < \log N +1.
\end{eqnarray}
Here, $\log$ denotes $\log_2$.
The total number of qubits is
$ 
L_1+L_2 \equiv L.
$ 
The state of each qubit is described by a vector 
of a Hilbert space spanned by two basis states, 
$| 0 \ket$ and $| 1 \ket$.
As a computational basis of register $R_{k}$, 
we take the set of the tensor products of the basis states of
$L_k$ qubits;
\begin{equation}
| a_1, a_2, \cdots, a_{L_k} \ket^{(k)}
\equiv
| a \ket^{(k)}.
\label{c-basis}\end{equation}
Here $a_\ell = 0$ or $1$ ($\ell=1,2,\cdots,L_k$) and
$ 
a \equiv \sum_{\ell = 1}^{L_k} a_\ell 2^{\ell -1},
$ 
for example, 
$|0 \ket^{(1)} =| 0 0 \cdots 0 \ket^{(1)}$,
$|1 \ket^{(1)} =| 1 0 \cdots 0 \ket^{(1)}$,
and 
$|2^{L_1}-1 \ket^{(1)} =| 1 1 \cdots 1 \ket^{(1)}$.
The initial state is taken as the following separable state;
\begin{equation}
|0 \rangle^{(1)} | 1 \rangle^{(2)}
\equiv 
|\psi_{\rm init} \rangle.
\label{psi0}\end{equation}
%
First, the Hadamard transformation is performed
by successive unitary transformations on individual 
qubits of $R_1$,
yielding 
\begin{equation}
\frac{1}{\sqrt{2^{L_1}}}\sum_{a=0}^{2^{L_1}-1}
|a \rangle^{(1)}
 |1 \rangle^{(2)}
\equiv |\psi_{\rm HT}\rangle.
\end{equation}
Then, we take randomly an integer ${\sf x}$ (${\sf x}<N$)
that is coprime to $N$ \cite{x}, and perform
the modular exponentiation
by successive pairwise unitary transformations,
yielding
\begin{equation}
\frac{1}{\sqrt{2^{L_1}}}\sum_{a=0}^{2^{L_1}-1}
|a \rangle^{(1)}
|{\sf x}^a \mbox{ mod } N\rangle^{(2)}
\equiv |\psi_{\rm ME}\rangle.
\label{psiME}\end{equation}
Finally, the discrete Fourier transformation (DFT) is performed
by successive pairwise unitary transformations,
yielding
\begin{equation}
\frac{1}{2^{L_1}} \sum_{a=0}^{2^{L_1}-1} \sum_{c=0}^{2^{L_1}-1}
\exp\left( \frac{2\pi i}{2^{L_1}} ca \right)
|\bar{c} \rangle^{(1)} |{\sf x}^a \mbox{ mod } N\rangle^{(2)}
\equiv |\psi_{\rm final}\rangle.
\label{psiDFT}\end{equation}
Here, $\bar{c}$ is the number that is obtained by reading the bits of 
$c$ in the reversed order \cite{Ekert_Jozsa}.
This completes the main part of Shor's factoring algorithm, 
and no more quantum computation is necessary.
Since one can read $\bar{c}$ reversely, 
the final state 
$|\psi_{\rm final}\rangle$ is practically equivalent to 
$
\frac{1}{2^{L_1}} \sum_{a=0}^{2^{L_1}-1} \sum_{c=0}^{2^{L_1}-1}
\exp\left[ \frac{2\pi i}{2^{L_1}} ca \right]
|c \rangle^{(1)} |{\sf x}^a \mbox{ mod } N\rangle^{(2)}.
$

The amplitude of $|\psi_{\rm final}\rangle$ has dominant peaks 
at basis states $|c \rangle^{(1)}$ satisfying
\begin{equation}
-\frac{r}{2} \leq
\bar{c} r \ {\rm mod} \ 2^{L_1}
\leq \frac{r}{2},
\label{barc_peak}\end{equation}
where $r$ ($<N$) 
is the ``order'' of ${\sf x} \mbox{ mod } N$
\cite{Shor_1,Ekert_Jozsa}.
By performing a measurement that diagonalizes
the computational basis $\{ |c \rangle^{(1)} \}$ of $R_1$, 
one can obtain a value of $c$ satisfying inequality (\ref{barc_peak})
with the probability greater than 
$4/\pi^2$.
When such an integer $c$ is obtained,
one can find uniquely for each $c$, using the continued fraction 
 expansion of $\bar{c}/2^{L_1}$, a fraction $c'/r$ that satisfies
\begin{equation}
\left| \frac{\bar{c}}{2^{L_1}} - \frac{c'}{r} \right|
\leq \frac{1}{2^{L_1+1}},
\label{c'}\end{equation}
where $c'$ is an integer.
If $c'$ happens to be coprime to $r$, one can know the value of $r$.
In other cases, where $c'$ is not coprime to $r$ or 
$c$ does not satisfy inequality (\ref{barc_peak}), 
one would obtain wrong results for $r$ or could not 
obtain $c'$ satisfying inequality (\ref{c'}).
Whether the obtained value of $r$ is correct or not can be checked easily
by calculating ${\sf x}^r$ mod $N$ using a classical computer.
When the correct value of $r$ is not obtained, 
one can perform the algorithm again.
It is known that 
one can successfully obtain the correct value
by repeating the algorithm $O(\log N)$ times.
When $r$ is thus obtained, 
one can know a factor of $N$ with the probability greater than $1/2$.
Therefore, one can factor $N$ efficiently by repeating the algorithm.

\section{Numerical simulation without noise}\label{sec-wo-noise}

In order to study conjecture (i), we perform
numerical simulations without noise in this section.

\subsection{Simplification and the number of computational steps}
\label{ss-cs}

The process of the modular exponentiation 
$|\psi_{\rm HT}\rangle \to |\psi_{\rm ME}\rangle$
costs $O(L_1^3)$ steps \cite{Shor_1,Ekert_Jozsa}.
Since it will turn out that $R_1$ takes a major role, 
we simplify operations on $R_2$ in our numerical simulations; namely, 
we represent the process $|\psi_{\rm HT}\rangle \to |\psi_{\rm ME}\rangle$
as the product of $L_1$ controlled unitary transformations 
$U^{a_\ell 2^{\ell-1}}$
($\ell=1, 2, \cdots, L_1$),
which is described in Box 5.2 of Ref.~\cite{Nielsen_Chaung}.
Since $U^{a_\ell 2^{\ell-1}}$ can be decomposed into $O(L_1^2)$ 
pairwise unitary transformations, the states appearing during 
the modular exponentiation in our simulations 
correspond to $L_1$ representative states out of $O(L_1^3)$ states.
As a result of this simplification, 
the total number $Q$ of computational steps, 
from $|\psi_{\rm init} \rangle$ to 
$|\psi_{\rm final}\rangle$, 
in our simulation becomes
\begin{equation}
Q = 2L_1 + \frac{L_1 (L_1+1)}{2}. 
\end{equation}
Here, $L_1$ comes from the Hadamard transformation and 
another $L_1$ from the modular exponentiation, 
whereas $L_1 (L_1+1)/2$ comes from the discrete Fourier transformation.
This $Q$ is smaller than $O(L_1^3)$ steps of a real 
computation because of the above simplification.

We denote 
the time interval between subsequent computational steps by $\tau$.
Although $\tau$ depends on the hardware of the quantum computer, 
this dependence does not matter in the following discussions because
we are only interested in the $N$ dependence, which is 
the only crucial factor in discussing the exponential speedup.
The total computational time, 
starting from $|\psi_{\rm init} \rangle$ and 
ending with the measurement of $|\psi_{\rm final}\rangle$, 
is given by
\begin{equation}
\tau_{\rm total} = (Q+1) \tau.
\label{tau}\end{equation}

\subsection{Choice of normalized additive operators}
\label{ss-choice}

To judge that a quantum state is {\em not} an AFS, fluctuations of {\em all} 
normalized additive operators have to be investigated. 
On the other hand, to judge that a state {\em is} an AFS, 
it is sufficient to find out {\em one} 
normalized additive operator $\hat{A}$ for which 
$\langle \Delta \hat{A}^{2} \rangle = O(L^{0})$.
In order to confirm conjecture (i), it is therefore sufficient
to find out {\em one} 
$\hat{A}$ for which a quantum state(s) appearing in the computational
process has an anomalously large fluctuation, 
$\langle \Delta \hat{A}^{2} \rangle = O(L^{0})$.

As normalized additive operators, 
we here consider the ``magnetization'' 
in three directions $\alpha=x,y,z$, 
for the total and the individual registers,
which are defined by 
\begin{equation}
\hat{M}_\alpha
=
\frac{1}{L}\sum_{\ell\in R_{1},R_{2}}
\hat \sigma_\alpha(\ell) 
\label{Ma}\end{equation}
and
\begin{equation}
{\hat{M}^{(k)}}_\alpha
=
\frac{1}{L_k}\sum_{\ell\in R_{k}}
\hat \sigma_\alpha(\ell),
\label{Mka}\end{equation}
respectively.
The maximum value of 
$\langle (\Delta \hat M_\alpha)^2 \rangle$ is unity, 
which is taken, e.g., by the following AFS:
\begin{equation}
\frac{1}{\sqrt{2}}|0\rangle^{(1)}|0\rangle^{(2)}
+
\frac{1}{\sqrt{2}}|2^{L_1}-1\rangle^{(1)}|2^{L_2}-1\rangle^{(2)}.
\end{equation}
For separable states,
on the other hand, 
$\langle (\Delta \hat M_\alpha)^2 \rangle$ becomes as small as
$ 
\langle (\Delta \hat M_\alpha)^2 \rangle \le 1/L.
$ 
This is consistent with the fact that any separable state is a NFS.
For example, $|\psi_{\rm init} \rangle$
is a separable state, for which 
$\langle (\Delta \hat{M}_\alpha)^{2}\rangle \le 1/L$ 
for all three directions $\alpha=x,y,z$.

\subsection{Anomalously fluctuating states are used in Shor's algorithm}
\label{ss-used}

We evaluate fluctuations of 
$\hat{M}_\alpha$ and ${\hat{M}^{(1)}}_\alpha$ for all 
states appearing in Shor's factoring algorithm,
for various values of $N$ and ${\sf x}$.
We assume in this section that noises are absent, because 
our purpose in this section is to find out AFSs.

\begin{figure}[htbp]
\begin{center}
\includegraphics[width=0.6\linewidth]{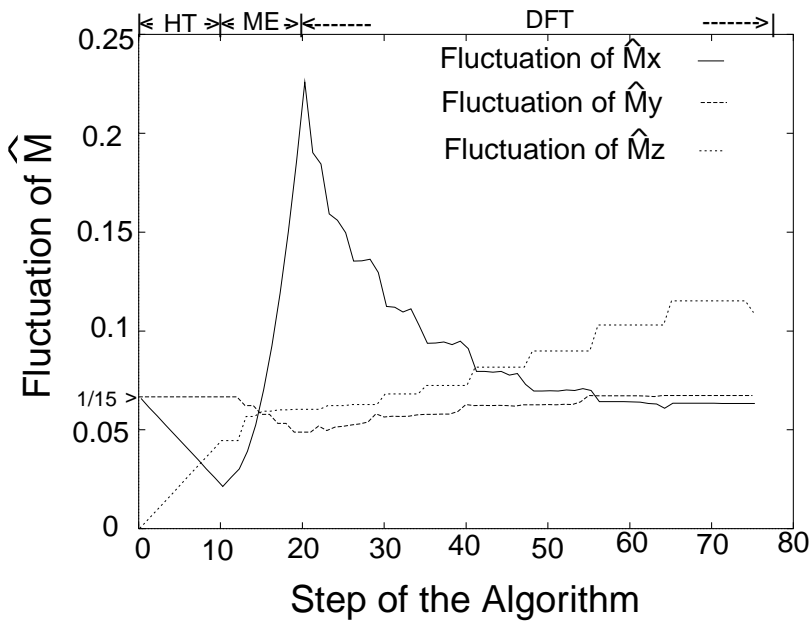}
\end{center}
\caption{$\langle(\Delta \hat M_\alpha)^2 \rangle$ 
of every state appearing in  Shor's factoring algorithm 
when
$N=21,\ L_1=10,\ L_2=5,\ {\sf x}=2$, for which $r=6$.
}
\label{yuragi15_1}
\end{figure}

Figure \ref{yuragi15_1} shows 
the change of 
$\langle(\Delta \hat M_\alpha)^2 \rangle$ 
along the steps of the algorithm when 
$N=21, L_1=10,\ L_2=5,\ {\sf x}=2$, for which $r=6$. 
It is seen that $\langle(\Delta \hat M_\alpha)^2\rangle$'s remain smaller than $1/L=1/15 \simeq 0.067$ until the Hadamard transformation is finished. 
This is consistent with the fact that any separable state is a NFS
(although the inverse is not necessarily true),
because all states during the Hadamard transformation are separable 
states [see, e.g., Eq.\ (\ref{HTx})].
In the modular exponentiation processes, 
on the other hand, 
$\langle(\Delta \hat M_x)^2 \rangle$ grows quickly, 
until it becomes $0.227$, 
which is significantly greater than $1/L$,
when the modular exponentiation is finished.
During the DFT, $\langle(\Delta \hat M_x)^2 \rangle$ decreases gradually, 
whereas $\langle(\Delta \hat M_z)^2 \rangle$ grows in turn 
until it becomes $0.109$, which is significantly
greater than $1/L$, at the final stage of the DFT.

\begin{figure}[htbp]
\begin{center}
\includegraphics[width=0.6\linewidth]{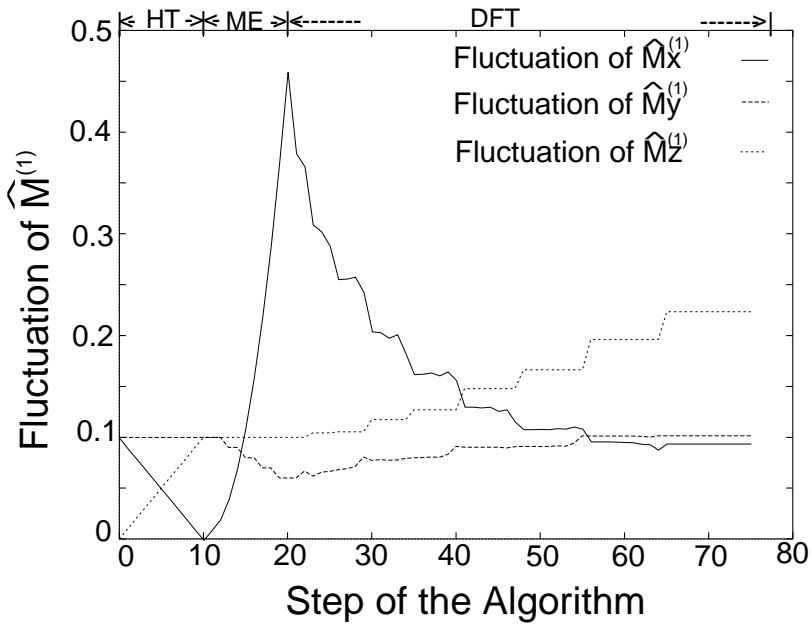}
\end{center}
\caption{
$\langle(\Delta \hat M_\alpha^{(1)})^2\rangle$ of every state 
appearing in Shor's factoring algorithm 
when
$N=21,\ L_1=10,\ L_2=5,\ {\sf x}=2$, for which $r=6$.
}
\label{yuragi15_2}
\end{figure}

To examine which register is responsible for 
the large fluctuations, 
we investigate $W_\alpha$ ($\alpha = x, y, z$) that is defined by 
\begin{equation}
W_\alpha
\equiv
\frac{\langle(\Delta \hat M_\alpha^{(1)})^2\rangle}{\langle(\Delta \hat M_\alpha)^2 \rangle}
=
\frac{
\langle (\Delta \hat M_\alpha^{(1)})^2\rangle
}{
\left\langle \left(
\frac{L_1}{L} \Delta {\hat{M}_\alpha}^{(1)} 
+ 
\frac{L_2}{L} \Delta {\hat{M}_\alpha}^{(2)}
\right)^{2}
\right\rangle}.
\end{equation}
If the first register gives dominant contribution to the fluctuation, we expect that
\begin{equation}
W_\alpha \simeq \biggl(\frac{L}{L_1}\biggr)^{2}=2.25.
\end{equation}
In Fig.\ \ref{yuragi15_2}, 
we plot the change of 
$\langle(\Delta \hat M_\alpha^{(1)})^2\rangle$
along the steps of the algorithm.
Comparing Fig.\ \ref{yuragi15_1} with Fig.\ \ref{yuragi15_2}, 
we find that $W_x = 2.03$ for $|\psi_{\rm ME}\rangle$ 
and $W_z = 2.05$ for $|\psi_{\rm final}\rangle$.
Since these values of $W_\alpha$ are close to $(L/L_1)^{2}$, 
we conclude that $R_1$ gives the dominant contribution.

\begin{figure}[htbp]
\begin{center}
\includegraphics[width=0.6\linewidth]{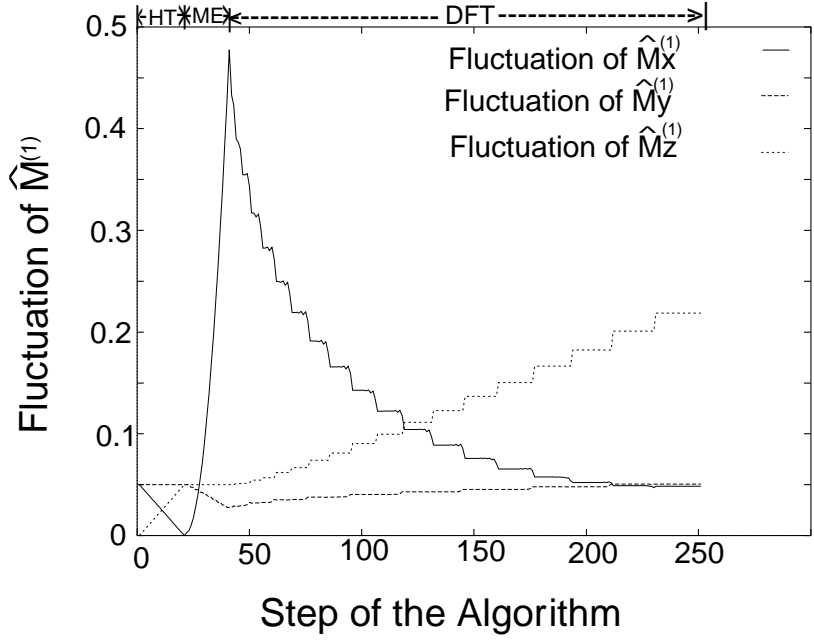}
\end{center}
\caption{
$\langle(\Delta \hat M_\alpha^{(1)})^2\rangle$ 
of every state appearing in  Shor's factoring algorithm 
when
$N=513,\ L_1=20,\ L_2=10,\ {\sf x}=26$, for which $r=6$.
}
\label{yuragi20_2}
\end{figure}

In order to judge whether the states are AFSs or not, 
we also investigate the $L$ 
dependence of the fluctuations.
Since $R_{1}$ gives dominant contributions, 
we calculate $\langle (\Delta \hat M_\alpha^{(1)})^2\rangle$ 
for a larger system with
$N=513,\ L_1=20,\ L_2=10,\ {\sf x}=26$, for which $r=6$.
Figure \ref{yuragi20_2} plots 
$\langle(\Delta \hat M_\alpha^{(1)})^2\rangle$ 
in this case.
By comparing this figure with Fig.\ \ref{yuragi15_2}, 
we find that 
$\langle (\Delta \hat M_x^{(1)})^2\rangle$ 
for $|\psi_{\rm ME}\rangle$ is almost independent of $L_1$.
In fact, 
$\langle\psi_{\rm ME}|(\Delta \hat M_x^{(1)})^2|\psi_{\rm ME}\rangle$ is 
0.460 and 0.477 in Figs.\ \ref{yuragi15_2} and \ref{yuragi20_2}, respectively.
We also find that 
$\langle (\Delta \hat M_z^{(1)})^2\rangle$ 
is almost independent of $L_1$
for $|\psi_{\rm final}\rangle$. 
In fact, $\langle\psi_{\rm final}|(\Delta \hat M_z^{(1)})^2
|\psi_{\rm final}\rangle$ 
is 0.223 and 0.219 in Figs.\ \ref{yuragi15_2} and \ref{yuragi20_2}, 
respectively.
On the other hand, 
other fluctuations have different dependences on $L_1$, 
for example, 
$\langle(\Delta \hat M_x^{(1)})^2\rangle$ for 
$|\psi_{\rm final}\rangle$ in Fig.\ \ref{yuragi20_2} is nearly 
half of that in Fig.\ \ref{yuragi15_2}.
Moreover, 
$\langle(\Delta \hat M_y^{(1)})^2\rangle$'s of all states  
in Fig.\ \ref{yuragi20_2} are nearly half of those
in Fig.\ \ref{yuragi15_2}. 

From these observations, 
we conclude that $|\psi_{\rm ME}\rangle$ and $|\psi_{\rm final}\rangle$ 
are AFSs, 
which have anomalously large fluctuations 
of $\hat M_x$ and $\hat M_z$, respectively.
Although this does not exclude the possibility that 
some other states could also be AFSs, 
identification of the above AFSs are sufficient for the 
purpose of the present paper; 
namely, we have confirmed conjecture (i)
when $(N, {\sf x}) = (21, 2)$ and $(513, 26)$,
both for which $r=6$.

We also performed numerical simulations for other values of 
$(N, {\sf x})$,
for which $r$ takes various values from $2$ to $20$.
We found that 
AFSs appear for all cases except when 
$r$ becomes an integral power of $2$, 
i.e., except when $r = 2, 4, 8, 16$; 
namely, we have confirmed conjecture (i) apart from 
the exceptional cases.
Since such exceptional cases seem to be unimportant in 
quantum computation, we consider a typical case $(N, {\sf x}) = (21, 2)$ 
in the following sections.

\subsection{Similarity of states with different values of $L$}
\label{ss-similarity}

From Figs.\ \ref{yuragi15_2} and  \ref{yuragi20_2}, 
it is seen that states of different values of $L$ have similar properties; 
namely, states in quantum computers
are homogeneous in a broad sense, because
states with a larger $L$ (which is necessary for larger inputs) 
and states with a smaller $L$
are both generated according to the same algorithm.
This allows us to analyze the $L$ dependence of 
the properties of quantum states in the quantum computers.


\section{Modeling and theoretical basis 
of quantum computers subject to noises}\label{modeling}

In order to examine conjectures (ii) and (iii), 
we investigate in the following sections 
effects of weak perturbations from classical noises 
on quantum computers.
In this section, we describe the modeling and theoretical basis.

\subsection{Model of noises}

We consider classical noises $f_x, f_y, f_z$, 
which act on the qubits through the following interaction Hamiltonian;
\begin{equation}
\hat H_{\rm int}
=
\sum_{\alpha = x, y, z} 
\lambda_\alpha
\sum_{\ell=1}^L
f_\alpha(\ell,t) \hat \sigma_\alpha(\ell).
\label{Hint}
\end{equation}
Since the computational basis is taken as Eq.\ (\ref{c-basis}), 
$f_{x}$ and $f_{y}$ are called ``bit-flip noises''
because they induce transitions between different basis states, 
whereas $f_{z}$ is called a ``phase-shift noise''
because
it induces phase shifts of the basis states.

In this paper, we consider long-wavelength noises, 
whose spectral intensities behave as Eq.\ (\ref{longwlnoise}); 
namely, we take
$f_\alpha(\ell,t)$ to be independent of $\ell$:
\begin{equation}
f_\alpha(\ell,t) = f_\alpha(t).
\label{fell}\end{equation}
For simplicity, we assume that
\begin{eqnarray}
&& \lambda_x = \lambda_y = \lambda_z \equiv \lambda,
\label{lambda}\\
&& \overline{f_\alpha(t)} = 0,
\label{avf}\\
&& 
\overline{f_\alpha(t) f_\beta(t')} 
= \delta_{\alpha, \beta} \
\overline{f_\alpha(t) f_\alpha(t')},
\label{avfafb}\end{eqnarray}
which seem natural in many physical situations.
Here, the overline denotes the average over 
the ensemble of the noises: 
\begin{equation}
\overline{\cdots}=\lim_{n_{\nu} \to \infty}
\frac{1}{n_{\nu}}\sum_{\nu=1}^{n_{\nu}}\cdots,
\label{avdot}\end{equation}
where $\nu$ labels realizations ($1, 2, \cdots, n_{\nu}$) of the noises. 
From Eqs.\ (\ref{fell}) and (\ref{lambda}), 
the interaction reduces to the simple form
\begin{equation}
\hat H_{\rm int}(t)
=
\lambda L
\sum_{\alpha = x, y, z}
f_\alpha(t) \hat M_\alpha.
\label{Hint_2}
\end{equation}

To simulate some real systems \cite{1fa,1fb,nakamura}, 
$f_\alpha(t)$ is assumed to have 
the $1/{\mathit f}$ spectrum \cite{1/f};
\begin{equation}
f_\alpha(t) = \sum_\omega  \frac{1}{\sqrt{\omega \tau}} 
\cos \left[\omega t + \theta_\alpha(\omega) \right],
\label{f(t)}\end{equation}
where $\theta_\alpha(\omega)$ is a random phase, 
which distributes uniformly in $(-\pi, \pi]$ for each $\omega$,
and the summation is taken over discrete values of $\omega$ 
in the interval 
$\omega_{\rm low} \leq \omega \leq \omega_{\rm high}$
with separations $\Delta \omega = 2 \pi/\tau_{\rm total}$.
Here, 
$\omega_{\rm low}$ and $\omega_{\rm high}$ are 
low- and high-frequency cutoffs, respectively.

In real systems, noises act continuously over the whole computational 
time $\tau_{\rm total}$.
However, we wish to study effects of noises 
{\em on each state} in the computation.
Therefore, we assume that the computation is perfectly performed 
until the $m$th step, and that the noises act between 
the $m$th and $(m+1)$th steps.
By calculating changes of the quantum states and 
of the computational results
in this case,
we can analyze effects of the noises on each state.

For this purpose, we take $\omega_{\rm low} = \Delta \omega$ because
the variation at such a low frequency is negligible
during the time interval $\tau$ of one step.
Regarding $\omega_{\rm high}$, 
we take 
$
\omega_{\rm high} \simeq 4.1 \times 2\pi/\tau
$, 
where the fractional factor $4.1$ is taken 
in order to 
avoid possible troubles which may occur by setting $\omega_{\rm high}$ as 
an integral multiple of $2\pi/\tau$.
We have confirmed that effects of higher-frequency components are negligible.

\subsection{Fidelity and decoherence}
\label{ss-fd}

Before presenting results of numerical simulations in the following section, 
we present in this section a theoretical basis for 
analyzing the numerical results.

As explained in Sec. \ref{ss-dr}, 
$\lambda$ can be interpreted as the 
noise amplitude times the coupling constant.
We are interested in the case of small $\lambda$, 
because otherwise it is obvious 
that the quantum computation would fail.
We therefore assume that $\lambda$ is small enough 
so that its 
lowest-order contribution is dominant
[see, e.g., Eq.\ (\ref{rho}) below].
Note, however, that the 
numerical results presented in the following section
include all orders in $\lambda$.
Since numerical simulations are possible only for relatively small $L$, 
we will draw general conclusions by using complementally both 
the numerical results and the analytic results of this section.

Suppose that the computation is perfectly performed 
until the $m$th step; namely, the state $|\psi_m \rangle$ 
that is obtained just after the 
$m$th step is exactly the state prescribed by the algorithm.
When the noises act between 
the $m$th and $(m+1)$th steps, 
the state evolves into 
\begin{eqnarray}
|\psi'_m \rangle
&=&
|\psi_m \rangle 
+\frac{1}{i \hbar} \int_{0}^{\tau} \! dt \ \hat H_{\rm int}(t) \, 
 |\psi_m \rangle
+
\frac{1}{(i \hbar)^2}
\int_{0}^{\tau} \! dt \ \int_{0}^{t} \! dt' \ 
\hat H_{\rm int}(t)
\hat H_{\rm int}(t')
|\psi_m \rangle
+\cdots.
\label{evolved}\end{eqnarray}
We are interested in the density operator $\hat \rho'_m$
that is the average of 
$| \psi'_m \rangle \langle \psi'_m |$ over the ensemble of the noises.
From Eqs.\ (\ref{avf}), (\ref{avfafb}), (\ref{Hint_2}), and 
(\ref{evolved}), 
we obtain 
\begin{eqnarray}
\hat \rho'_m
&\equiv& \overline{| \psi'_m \rangle \langle \psi'_m |}
\nonumber\\
&=& 
\hat \rho_m
-
\frac{\lambda^{2} L^{2}}{2 \hbar^2} \sum_\alpha C_\alpha(\tau) 
\left( 
\hat M_\alpha^2 \hat \rho_m 
+ \hat \rho_m \hat M_\alpha^2 
-2 \hat M_\alpha \hat \rho_m \hat \rho_m \hat M_\alpha 
\right) 
+ O(\lambda^3).
\label{rho}
\end{eqnarray}
Here, 
$
\hat \rho_m
\equiv
|\psi_m \rangle \langle \psi_m|
$, 
and
$C_\alpha(\tau)$ is an integral of the
autocorrelation function of the noise,
\begin{equation}
C_\alpha(\tau)
\equiv
2 \int_{0}^{\tau} \! dt \ \int_{0}^{t} \! dt' \ 
\overline{f_\alpha(t)f_\alpha(t')}
=
\int_{0}^{\tau} \! dt \ \int_{0}^{\tau} \! dt' \ 
\overline{f_\alpha(t)f_\alpha(t')}
=
\overline{
\left(
\int_{0}^{\tau} \! f_\alpha(t) dt 
\right)^2
}.
\end{equation}

As a measure of deviation of $\hat \rho'_m$ from 
the ideal one $\hat \rho_m$, 
we consider the fidelity that is defined by
\begin{equation}
F_m \equiv \mbox{Tr}\left[ \rho_m \rho_m' \right].
\label{F}\end{equation}
This is not necessarily a good measure of decoherence, because
it can be reduced even when 
$\hat \rho'_m$ happens to be a pure state. 
As a measure of decoherence, we consider 
the $\alpha$ entropy of $\alpha=2$,
\begin{equation}
S_m \equiv -\ln(\mbox{Tr}[(\rho'_m)^{2}]).
\label{S_m}\end{equation}
To see the relation between $F_m$ and $S_m$, 
let us represent Eq.\ (\ref{rho}) as
$ 
\hat \rho_m' 
=
\hat \rho_m + \lambda^2 \hat \eta_m +O(\lambda^{3}).
$ 
Then we obtain
\begin{eqnarray}
S_m 
&=& 
-2\lambda^{2}\mbox{Tr}[\hat \rho_m \hat \eta_m] +O(\lambda^{3}),
\\
F_m
&=&
1+
\lambda^{2}\mbox{Tr}[\hat \rho_m \hat \eta_m] +O(\lambda^{3}).
\end{eqnarray}
We thus find
\begin{equation}
S_m = 2(1-F_m)+O(\lambda^{3}).
\label{SofF}\end{equation}
Therefore, to $O(\lambda^2)$, 
the reduction of the fidelity is directly related to the increase 
of the entropy, i.e., to the decoherence.
Since we are interested in the case of small $\lambda$,
this perturbative relation should hold.
We thus simply call the reduction of the fidelity 
``decoherence'' in the following discussions.

Inserting Eq.\ (\ref{rho}) into Eq.\ (\ref{F}), 
we obtain a simple formula, 
which is correct up to $O(\lambda^2)$,
\begin{eqnarray}
F_m 
&=&
1-
\frac{\lambda^2 L^2}{\hbar^2} \sum_\alpha 
C_\alpha (\tau)
\langle \psi_m | (\Delta \hat M_\alpha)^2 | \psi_m \rangle,
\label{F_tau}
\end{eqnarray}
where
$ 
\Delta{\hat{M}_\alpha}
\equiv
\hat{M}_\alpha
-
\langle \psi_m | \hat M_\alpha | \psi_m \rangle.
$ 
It is seen that, to $O(\lambda^2)$,
$F_m$ decreases in proportion to 
$L^2$ and to the fluctuations of magnetizations, 
in accordance with the general result, Eq.\ (\ref{r-noise-2}).
It is also seen that 
the noises in three directions $f_x, f_y, f_z$
contribute additively to $F_m$.
We can therefore calculate effects of $f_x, f_y, f_z$ independently.

The decoherence rate (per step)
$\Gamma_m$ of $| \psi_m \rangle$ is given by 
\begin{equation}
\Gamma_m \equiv
\frac{S_m}{2 \tau},
\label{Gamma_m}\end{equation}
where the factor of 2 has been inserted for convenience.
From Eqs.\ (\ref{SofF}) and (\ref{F_tau}),
we obtain a formula for $\Gamma_m$;
\begin{equation}
\Gamma_m
=
\frac{\lambda^2 L^2}{\hbar^2} \sum_\alpha 
\frac{C_\alpha (\tau)}{\tau}
\langle \psi_m | (\Delta \hat M_\alpha)^2 | \psi_m \rangle,
\label{Gamma}
\end{equation}
which is correct up to $O(\lambda^2)$.
This formula is a special case of
the general result of SM, 
Eq.\ (\ref{r-noise-2}), 
except for the extra factor $C_\alpha (\tau)/\tau$.
The case of $\alpha = z$ of the above formula
agrees also with 
the result of Ref.\ \cite{Palma}, 
except for the extra factor.
The extra factor appears because 
the $1/\mathit{f}$ noise, Eq.\ (\ref{f(t)}), does not satisfy 
the assumption of short-time correlation.
Since the extra factor is independent of the state of the qubit system,
it can be considered as a constant when comparing decoherence rates
of different states.

\subsection{Fragility of anomalously fluctuating states}

When $| \psi_m \rangle$ is a NFS, 
$
\langle \psi_m | (\Delta \hat M_\alpha)^2 | \psi_m \rangle
\leq O(1/L)
$
by definition.
Formula (\ref{Gamma}) then yields
\begin{equation}
\Gamma_m \leq \lambda^2 O(L).
\label{G-NFS}\end{equation}
Therefore, NFSs never become fragile.
When $| \psi_m \rangle$ is an AFS that has been found in 
Sec \ref{ss-used},
on the other hand, 
$
\langle \psi_m | (\Delta \hat M_\alpha)^2 | \psi_m \rangle
= O(L^0)
$
for $\alpha=x$ or $z$.
If the noise component $f_\alpha$ for such $\alpha$ is present,
formula (\ref{Gamma}) yields
\begin{equation}
\Gamma_m = \lambda^2 O(L^2),
\label{G-AFS}\end{equation}
hence the AFS becomes fragile (with $\delta = 1$).

These results are 
consistent with conjecture (ii) and with
the more general results that are summarized in 
Sec. \ref{ss-frg-of-AFS}.
It is important to note the following fact, which is obtained 
from Eqs.\ (\ref{G-NFS}) and (\ref{G-AFS}):
\begin{equation}
\frac{\mbox{(decoherence rate of fragile AFSs)}}
{\mbox{(decoherence rate of NFSs)}}
\geq O(L).
\end{equation}
This ratio becomes $\gg 1$ when $L \gg 1$, i.e., when the input 
$N$ is huge.
Therefore, the decoherence rate of the quantum computer is 
almost determined by the decoherence rates of the fragile AFSs.
These points will be studied more in detail using numerical 
simulations in the following section.

Note also that decoherence does not necessarily reduce the 
success probability
of the computation.
We will also study this point in the following sections.

\section{Results of numerical simulations with noises}\label{sec-wt-noise}

In this section, we present 
results of numerical simulations of Shor's factoring algorithm
when noises act on a quantum computer.
As explained in subsection \ref{ss-used}, 
we take 
$N=21,\ L_1=10,\ L_2=5,\ {\sf x}=2$, for which $r=6$ and $Q=75$.

\subsection{Decoherence} 

\begin{figure}[htbp]
\begin{center}
\includegraphics[width=0.6\linewidth]{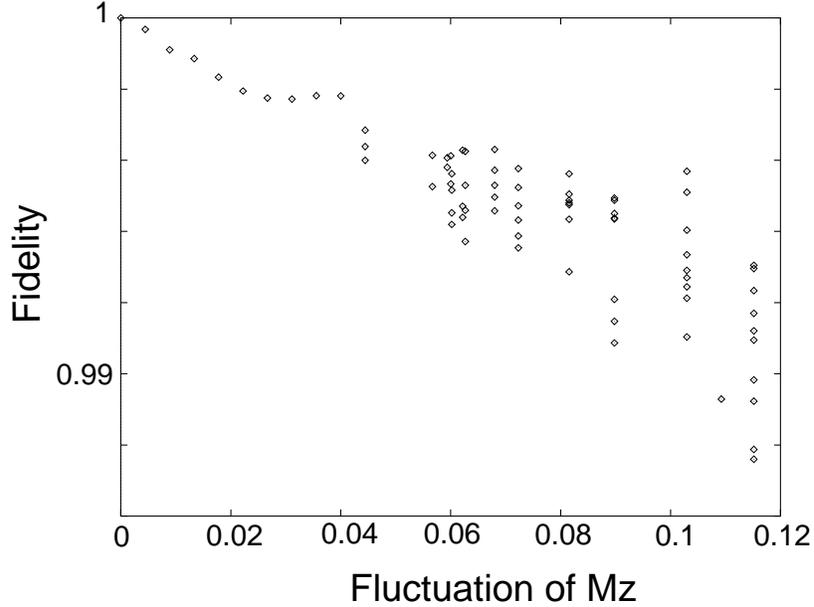}
\end{center}
\caption{
Fidelity vs $\langle (\Delta \hat M_z)^2  \rangle$
in the presence of a phase-shift noise $f_z$ 
when $N=21,\ {\sf x}=2$, and $\lambda=0.0015 \hbar/\tau$.
The average over noise realizations has been taken over 
$40$ samples.
}
\label{Mz_Dec}
\end{figure}

\begin{figure}[htbp]
\begin{center}
\includegraphics[width=0.6\linewidth]{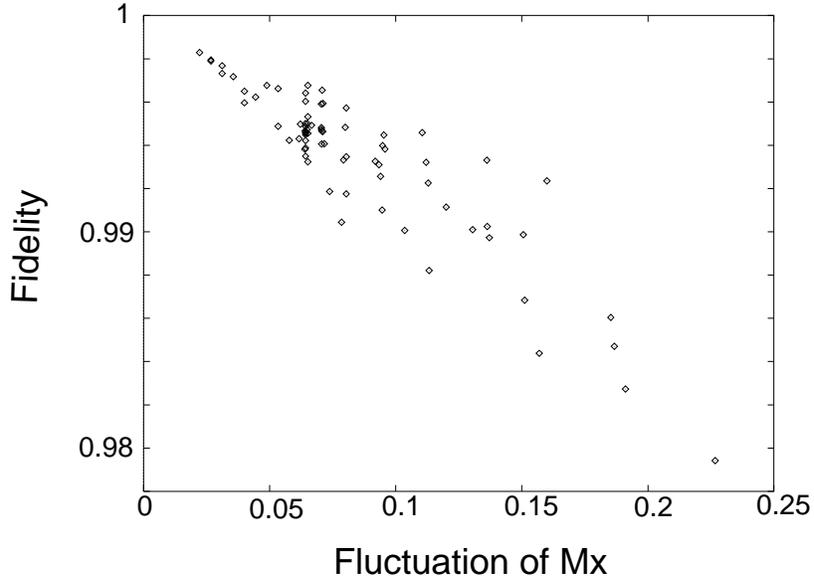}
\end{center}
\caption{
Fidelity vs $\langle (\Delta \hat M_x)^2  \rangle$
in the presence of a bit-flip noise $f_x$ 
when $N=21,\ {\sf x}=2$, and $\lambda=0.0015 \hbar/\tau$.
The average over noise realizations has been taken over 
$40$ samples.
}
\label{Mx_Dec}
\end{figure}

In Fig.\ \ref{Mz_Dec},  we show 
a two-dimensional plot of
the fidelity versus the fluctuation of $\hat{M}_z$ for all steps, 
i.e., 
($\langle \psi_m | (\Delta \hat M_z)^2 | \psi_m \rangle$, $F_m$)
for all $m$, 
when only a phase-shift noise is present, 
i.e., when $f_z$ is given by Eq.\ (\ref{f(t)}) whereas 
$f_x = f_y =0$.
Each point in the figure corresponds to a state appearing in the algorithm. 
The parameter $\lambda$ is taken as $0.0015 \hbar/\tau$.
Regarding a bit-flip noise, 
we show in Fig.\ \ref{Mx_Dec} a two-dimensional plot of
the fidelity versus the fluctuation of $\hat{M}_{x}$ 
when $f_x$ is given by Eq.\ (\ref{f(t)}) whereas 
$f_y = f_z =0$ for $\lambda=0.0015 \hbar/\tau$.
A similar plot (not shown) has been obtained if we plot 
($\langle \psi_m | (\Delta \hat M_y)^2 | \psi_m \rangle$, $F_m$)
when $f_y$ is given by Eq.\ (\ref{f(t)}) whereas
$f_x=f_z=0$.
This is reasonable because both $f_x$ and $f_y$ are bit-flip noises.

Figures \ref{Mz_Dec} and \ref{Mx_Dec}
show that $F$ decreases, on an average, in proportion to 
the fluctuation of the normalized additive operator to 
which the noise couples via 
$\hat H_{\rm int}(t)$ of Eq.\ (\ref{Hint_2}).
Distributions around the average curve are due to 
the fact that the number $n_\nu$ of noise samples 
is not very large; $n_\nu = 40$. 
Therefore, 
formula (\ref{F_tau}) has been confirmed.
Hence, formulas (\ref{Gamma})-(\ref{G-AFS}) have also been confirmed.

In these figures, states with larger fluctuations are more likely 
to be AFSs.
In particular, the states with the largest fluctuations
in Figs.\ \ref{Mz_Dec} and \ref{Mx_Dec} are
$|\psi_{\rm final}\rangle$ and 
$|\psi_{\rm ME}\rangle$, 
respectively, 
which have been identified as AFSs in Sec. \ref{ss-used}.
Although $L$ is rather small ($L=15$) in this simulation, 
we can extrapolate the results to the case of larger $L$
using formulas (\ref{Gamma})-(\ref{G-AFS}).
(See discussions in subsection \ref{ss-crucailAFS}.)
It is then clear that conjecture (ii) is correct.

\subsection{Success probability}
\label{ss-sp}

Generally speaking, 
decoherence of a particular state does not necessarily lead to 
false results of the quantum computation.
For example, as will be discussed in Sec. \ref{irrelAFS}, 
decoherence of $| \psi_{\rm final} \rangle$ by
the phase-shift noise $f_z$ does not reduce the probability of
getting the correct value of $r$ at all.
Therefore, in this section we investigate effects of the noises on 
the computational result for each state.

\begin{figure}[htbp]
\begin{center}
\includegraphics[width=0.6\linewidth]{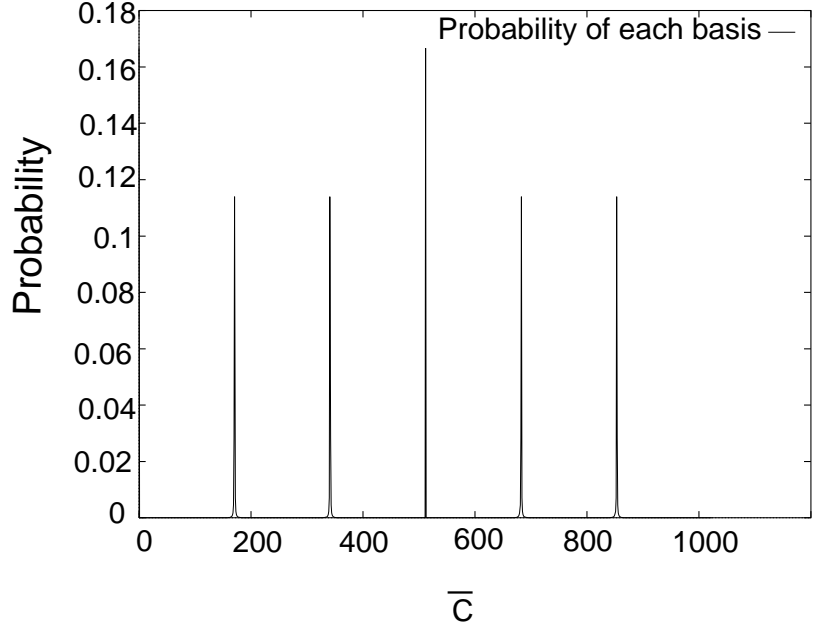}
\end{center}
\caption{Probability distribution of finding a basis state 
$| c \rangle^{(1)}$ at the end of the algorithm 
when $N=21,\ {\sf x}=2$, and $\lambda=0$.
The horizontal axis is the bit reversal $\bar{c}$ of $c$. 
}
\label{peak}
\end{figure}

\begin{figure}[htbp]
\begin{center}
\includegraphics[width=0.6\linewidth]{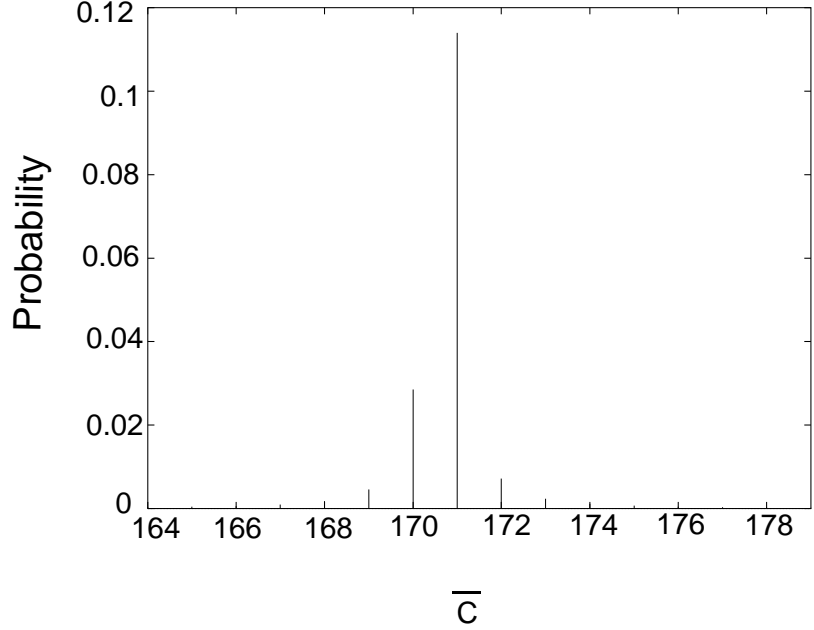}
\end{center}
\caption{
A magnification of Fig.\ \ref{peak} around a dominant peak at
$\bar{c} = 171$.
}
\label{peak_2}
\end{figure}

For this purpose, we 
calculate the {\it success probability} $T$ 
that is defined as the probability of 
finding in the final state a basis state 
which gives the correct value of $r$.
In Shor's factoring algorithm, 
one performs measurement, which diagonalize the computational basis, 
on the final state $| \psi_{\rm final} \rangle$.
When noises are absent, for example, 
we plot in Fig.\ \ref{peak} the probability of finding each 
basis state in $| \psi_{\rm final} \rangle$.
Each dominant peak is accompanied by side peaks, 
as shown in Fig.\ \ref{peak_2}, which is a magnification of Fig.\ \ref{peak}.
If the basis state with $c = \overline{171}$ or $\overline{853}$ 
happens to be obtained,  
one can successfully obtain the correct value of $r=6$, where the overline 
denotes the bit reversal.
The other basis states do not give the correct value.
For example, if one obtains $c = \overline{512}$, which corresponds to 
the central peak in Fig.\ \ref{peak}, 
then $c'$ satisfying inequality (\ref{c'}) is $c' = 3$.
Since $c'$ is not coprime to $r=6$, it gives a wrong result as
$r=2$.
Whether the result is correct or not
can be checked efficiently using a classical computer.
Furthermore, if one obtains $c = \overline{170}$, which corresponds to 
the highest side peak associated with the dominant peak at 
$c = \overline{171}$ 
in Fig.\ \ref{peak_2}, 
then there is no $c'$ satisfying inequality (\ref{c'}).
Therefore, $T$ is given in the case of $N=21,\ {\sf x}=2$ by
\begin{equation}
T =
P(\overline{171}) + P(\overline{853}),
\label{T}\end{equation}
where $P(c)$ denotes the probability of finding in the final state a 
basis state 
$| c \rangle^{(1)}$.

It is essential to exclude $P(c)$'s of the basis 
states corresponding to 
the other peaks.
In fact, we found that some of such excluded $P(c)$'s 
sometimes {\em increase} with increasing the strength of noises.
Hence, we would have obtained unphysical results if we included them
in the success probability.

It is evident from Eq.\ (\ref{T}) that 
$T < 1$ even in the absence of the noises. 
If we denote $T$ in the absence of noises by $T_{\rm clean}$, 
it is evaluated for the case of Fig.\ \ref{peak} as
$T_{\rm clean} = 0.22797$.
When noises act on the quantum computer, $T$ would become smaller than 
$T_{\rm clean}$,
\begin{equation}
T = \epsilon T_{\rm clean},
\label{epsilon}\end{equation}
where $\epsilon \leq 1$.
For a successful quantum computation, 
the factor $\epsilon$ should be kept larger than 
some threshold value $\epsilon_{\rm th}$ (see 
Sec. \ref{ss-sconl}):
\begin{equation}
\epsilon \geq \epsilon_{\rm th}.
\end{equation}

Consider now the case where
noises act {\em only} between 
the $m$th and $(m+1)$th steps.
We denote the success probability in this case by $T_m$.
By investigating $T_m$ and $F_m$ for each $m$, 
we can study effects of the noises on 
the computational results for each state $|\psi_m \rangle$.
In real situations, noises act continuously throughout all steps.
Hence, the success probability in real situations
is less than any $T_m$:
\begin{equation}
T \leq \min_m T_m.
\label{Treal}\end{equation}
Therefore, the following condition is necessary 
for a successful quantum computation:
\begin{equation}
\min_m \frac{T_m}{T_{\rm clean}} \geq \epsilon_{\rm th}.
\label{Tth}\end{equation}

\subsection{Relation between the success probability and decoherence}

\begin{figure}[htbp]
\begin{center}
\includegraphics[width=0.6\linewidth]{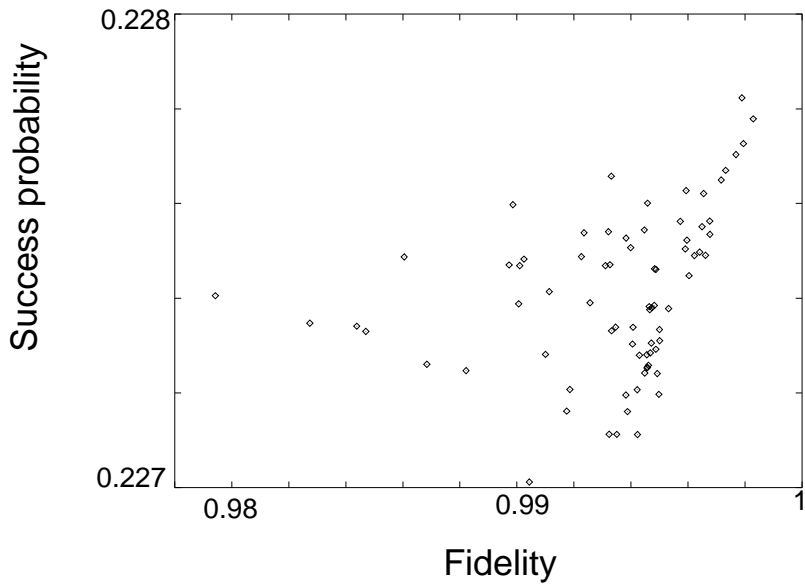}
\end{center}
\caption{
The success probability vs the fidelity in the 
presence of a bit-flip noise $f_x$ 
when $N=21,\ {\sf x}=2$, and $\lambda=0.0015 \hbar/\tau$.
The average over noise realizations has been taken over 
$40$ samples.
}
\label{Decxtot}
\end{figure}

\begin{figure}[htbp]
\begin{center}
\includegraphics[width=0.6\linewidth]{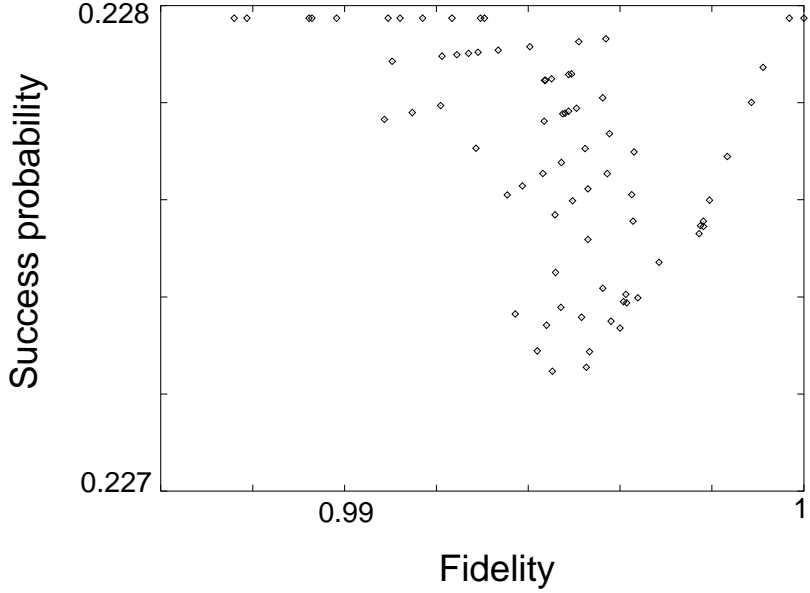}
\end{center}
\caption{
The success probability vs the fidelity in the 
presence of a phase-shift noise $f_z$ 
when $N=21,\ {\sf x}=2$, and $\lambda=0.0015 \hbar/\tau$.
The average over noise realizations has been taken over 
$40$ samples.
}
\label{Decztot}
\end{figure}

In Figs.\ \ref{Decxtot} and \ref{Decztot},
we plot the relation between the fidelity $F_m$ 
and the success probability $T_m$ when $\lambda=0.0015 \hbar/\tau$ 
for a bit-flip noise [$f_x \neq 0$, given by Eq.\ (\ref{f(t)}], 
whereas $f_y = f_z = 0$) 
and a phase-shift noise [$f_x = f_y = 0$, 
whereas $f_z \neq 0$, given by Eq.\ (\ref{f(t)})], respectively;
namely, $(F_m, T_m)$ are plotted for all $m$.

It is seen that for most states their decoherence results 
in the reduction of the success probability, as might be expected.
If we define the {\em rate of the 
reduction of the success probability per decoherence} by
\begin{equation}
r_m \equiv
\frac{T_{\rm clean}-T_m}{1-F_m},
\label{rm}\end{equation}
then it is found that $r_m$ varies from state to state.

We notice that there are exceptional states 
located at the top of Fig.\ \ref{Decztot}, 
for which the phase-shift noise does not reduce the success probability.
These are states which appear in the latter part of the DFT.
This means that quantum coherence among 
basis states of these states is unnecessary for quantum computation.
This may be understood by considering effects of the phase-shift noise 
on the final state $|\psi_{\rm final} \rangle$.
In order to estimate $r$, 
one will perform measurement on this state, 
which diagonalizes the computational basis.
This means that one will not measure the relative phases among 
basis states.
Therefore, decoherence of the relative 
phases in $|\psi_{\rm final} \rangle$ 
does not change the computational result at all. 
This point will be discussed later again in Sec. \ref{irrelAFS}.

\subsection{An AFS is crucial}
\label{ss-crucailAFS}

We now investigate differences between 
effects of noises on NFSs and those on AFSs.
For this purpose, we investigate effects of noises 
on $|\psi_{\rm HT}\rangle$, which is a typical NFS,
and those on $|\psi_{\rm ME}\rangle$, which is a typical AFS. 

\begin{figure}[htbp]
\begin{center}
\includegraphics[width=0.6\linewidth]{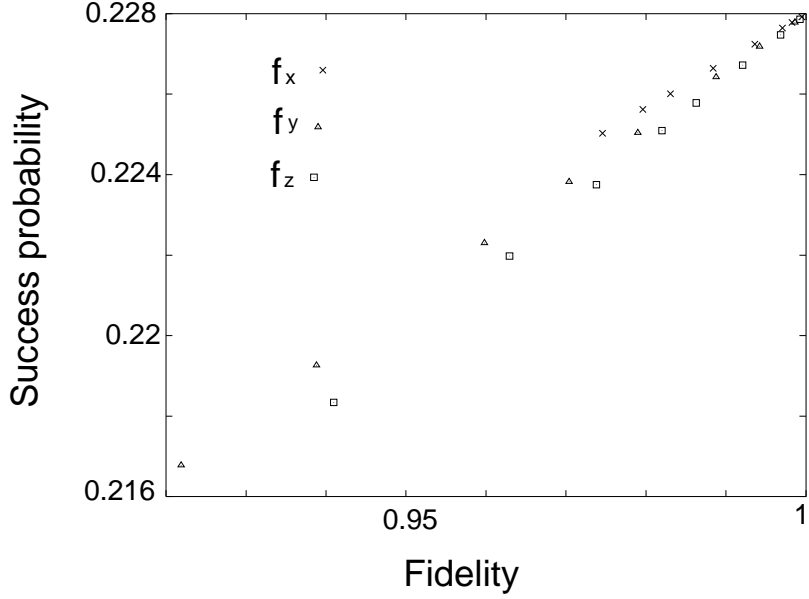}
\end{center}
\caption{
The success probability vs the fidelity of
$|\psi_{\rm HT}\rangle$, which is a typical NFS, 
for various strength of the noises $f_x, f_y, f_z$.
The average over noise realizations has been taken over 
$200$ samples.
}
\label{NOISEonHT}
\end{figure}

\begin{figure}[htbp]
\begin{center}
\includegraphics[width=0.6\linewidth]{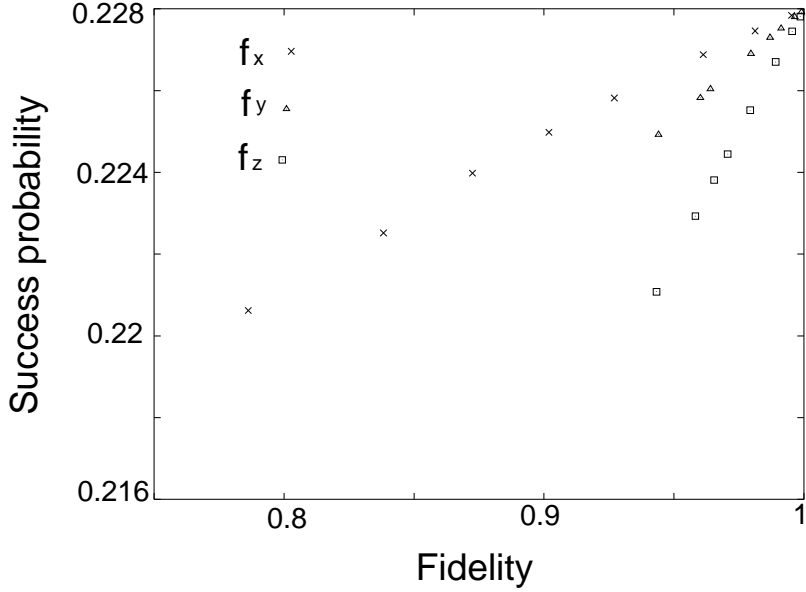}
\end{center}
\caption{
The success probability vs the fidelity of
$|\psi_{\rm ME}\rangle$, which is a typical AFS, 
for various strength of the noises $f_x, f_y, f_z$.
The average over noise realizations has been taken over 
$200$ samples.
}
\label{NOISEonME}
\end{figure}

Figures \ref{NOISEonHT} and \ref{NOISEonME} plot $(F_m, T_m)$ 
for 
$| \psi_m \rangle = |\psi_{\rm HT}\rangle$ 
and for 
$| \psi_m \rangle = |\psi_{\rm ME}\rangle$, 
respectively, 
for various values of $\lambda$ ranging from 
$0.00075 \hbar/\tau$ to $0.006 \hbar/\tau$.
The crosses represent the case of a bit-flip noise with
$f_x \neq 0$, given by Eq.\ (\ref{f(t)}), whereas $f_y = f_z = 0$.
The triangles represent the case of another bit-flip noise with
$f_y \neq 0$, given by Eq.\ (\ref{f(t)}), whereas $f_x = f_z = 0$.
On the other hand, 
the squares in the figures represent the case of a phase-shift noise with
$f_z \neq 0$, given by Eq.\ (\ref{f(t)}), 
whereas $f_x = f_y = 0$.

It is found that 
both $F_m$ and $T_m$ decrease with increasing $\lambda$.
As a result, $(F_m, T_m)$ of either state 
composes a monotonic curve, 
although the data scatter slightly around the average curve because 
the average over noise realizations has been taken over a finite 
number ($n_\nu = 200$) of samples.
The average curves are almost straight 
for such small $\lambda$
as assumed in the numerical simulations;
namely, for small $\lambda$, 
\begin{equation}
r_m \simeq 
\mbox{ const,\ independent of $\lambda$, 
for each noise and $m$},
\label{rmisconst}\end{equation}
where $r_m$ is defined by Eq.\ (\ref{rm}).

For $|\psi_{\rm HT}\rangle$, the rate 
$r_m$ takes similar values for the three noises.
Regarding each of $F_m$ and $T_m$, 
on the other hand, the magnitudes of their reductions
by $f_x$ are smaller than those by $f_y$ or $f_z$.
This may be understood by noting that 
$|\psi_{\rm HT}\rangle$ can be rewritten simply as
\begin{equation}
|\psi_{\rm HT}\rangle
=
|+x, +x, \cdots, +x \rangle^{(1)}
 |1 \rangle^{(2)},
\label{HTx}\end{equation}
where $|+x, +x, \cdots, +x \rangle^{(1)}$ denotes 
the simultaneous eigenstate of 
$\hat \sigma_x(1), \hat \sigma_x(2), \cdots, \hat \sigma_x(L_1)$
with the eigenvalues $+1, +1, \cdots, +1$.
It is then clear that 
the state of $R_1$ is not changed by $f_x$ which couples
to $\hat \sigma_x(\ell)$ through the interaction (\ref{Hint}),
because $f_x$ induces only a random phase factor $e^{i \theta}$ as
$ 
|+x, +x, \cdots, +x \rangle^{(1)}
\to
e^{i \theta} |+x, +x, \cdots, +x \rangle^{(1)},
$ 
which means no change in the quantum state.
Hence, the reductions of $F_m$ and $T_m$ of $|\psi_{\rm HT}\rangle$
occur only through 
the decoherence of the state of $R_2$, 
and thus the reductions become small.

For $|\psi_{\rm ME}\rangle$, 
the reduction of $F_m$ by $f_x$ 
is larger than that 
by $f_y$ or $f_z$.
This can be understood from Eq.\ (\ref{F_tau}):
Through the interaction (\ref{Hint_2}), 
$f_x$ couples to $\hat M_x$, 
which has an anomalously large fluctuation for this state, 
whereas 
$f_y$ and $f_z$ couple to $\hat M_y$ and $\hat M_z$, respectively, 
which have normal fluctuations.
Therefore, it is confirmed again that 
the reduction of $F_m$ is simply determined by the 
fluctuation of $\hat M_\alpha$ to which the noise couples.
Regarding $T_m$, on the other hand, 
there is no significant difference between 
the magnitudes of the reductions by $f_x$ and $f_z$, 
whereas the reduction by $f_y$ is smaller.
As a result,  $r_m$ takes different values for the three noises.

It seems reasonable to assume that in real systems
noises in all directions coexist with similar magnitudes, 
because noises are uncontrollable, 
random objects.
Then, it may be tempting from Figs.\ \ref{NOISEonHT} and \ref{NOISEonME}
to say that the 
effects on the computational result 
of the noises acting on $|\psi_{\rm HT}\rangle$
and those acting on $|\psi_{\rm ME}\rangle$
would be of the same order of magnitudes.
However, to study the performance of computations, 
we must consider how the effects of noises scale as $L$ is increased.
Although $L$ is not so large ($L=15$) in Figs.\ \ref{NOISEonHT} and
\ref{NOISEonME}, 
it will be much larger in practical applications.
In order to satisfy condition (\ref{Tth}) for larger $L$, 
$\lambda$ should be made smaller.
Suppose that one somehow reduces $\lambda$ with increasing $L$
in such a way that 
$1-F_m$ for $|\psi_{\rm HT}\rangle$ is kept constant.
According to Eq. (\ref{G-NFS}), 
this means the following scaling rule:
\begin{equation}
\lambda^2 \propto 1/L.
\label{scaleNFS}\end{equation}
However, 
according to Eq.\ (\ref{G-AFS}), 
$1-F_m$ for $|\psi_{\rm ME}\rangle$ scales in this case as
$1-F_m \propto L$.
Then, 
since $T_m$ is a monotonically decreasing function with decreasing $F_m$, as 
seen from Fig.\ \ref{NOISEonME} and Eq.\ (\ref{rmisconst}) \cite{nonlinear}, 
it is expected that 
$T_m$ is greatly reduced, approaching zero, with increasing $L$.
Therefore, condition (\ref{Tth}) would be violated
in practical applications, for which $L$ is large, if $\lambda$ is 
scaled as Eq.\ (\ref{scaleNFS}).

This argument is applicable also to other AFSs and NFSs that appear 
during quantum computations. 
Therefore, $\displaystyle \min_m T_m$ 
on the left-hand side of condition (\ref{Tth})
is $T_m$ of an AFS, and 
it is necessary that 
\begin{equation}
\lambda^2 \leq O(1/L^2),
\label{conditon}\end{equation}
which is much severer than the naive one (\ref{scaleNFS}).
This supports conjecture (iii); namely, we conclude that the bottleneck of 
the quantum computations is the anomalously fast 
decoherence of AFSs, which appear 
during quantum computations.

Note that inequality (\ref{conditon}) is {\it not} a sufficient condition, 
but a necessary condition.
We will discuss a sufficient condition in Sec. \ref{sec-discussion}.

\subsection{Some AFSs may be noncrucial}
\label{irrelAFS}

We have shown that the decoherence of 
{\em one} of AFSs appearing in Shor's factoring algorithm
is crucial, in consistency with conjecture (iii).
Note however that this does {\em not} mean that decoherence of 
{\em all} AFSs appearing in the algorithm were crucial.

For example, consider the final state $|\psi_{\rm final} \rangle$, 
which has an 
anomalously large fluctuation
of $\hat M_z$, 
whereas the fluctuations of $\hat M_x$ and $\hat M_y$ take
normal values. 
When a phase-shift noise with $f_z \neq 0, f_x=f_y=0$ acts
on this state, it evolves into 
\begin{equation}
|\psi'_{\rm final}\rangle
=
\frac{1}{2^{L_1}} \sum_{a=0}^{2^{L_1}-1} \sum_{c=0}^{2^{L_1}-1}
\exp\left( \frac{2\pi i}{2^{L_1}} ca + i \theta(c,a)\right)
|\bar{c} \rangle^{(1)} |{\sf x}^a \mbox{ mod } N\rangle^{(2)},
\end{equation}
where $\theta(c,a)$ is a phase, which 
depends on $c, a,$ and the noise.
Since the probability $P(c)$ of finding a base 
$|c \rangle^{(1)}$ does not depend on this random phase,
$T_m$ for this state is not reduced at all by
the phase-shift noise, 
whereas the fidelity $F_m$ is reduced in proportion to $L^2$.
Therefore, unlike $|\psi_{\rm ME}\rangle$, 
the anomalously fast decoherence of 
$|\psi_{\rm final} \rangle$
is not crucial. 
$T_m$ of this state is reduced only by $f_x$ or $f_y$, 
in proportion to $L$.


\section{discussions}
\label{sec-discussion}

\subsection{A sufficient condition on $\lambda$}
\label{ss-sconl}

Inequality (\ref{conditon}) is not a sufficient condition, 
but a necessary condition.
In this section, we discuss a sufficient condition.

In real situations, 
noises act continuously throughout the computation,
as discussed in Sec. \ref{ss-sp}.
Hence, the success probability $T$ in real situations may be given by
\begin{eqnarray}
T
&\simeq&
T_{\rm clean} \prod_m \frac{T_m}{T_{\rm clean}}
\end{eqnarray}
For a sufficiently small $\lambda$, 
since $T_{\rm clean} - T_m \propto \lambda^2$, 
this may be approximated by
\begin{eqnarray}
T
&\simeq&
T_{\rm clean} \exp 
\left(- \sum_m \frac{T_{\rm clean} - T_m}{T_{\rm clean}}\right)
\nonumber\\
&=&
T_{\rm clean} \exp 
\left[
- \lambda^2 \left(
\sum_{m \ \in \mbox{ crucial AFSs}} O(L^2)
+\sum_{m \ \in \mbox{ noncrucial AFSs and NFSs}} O(L)
\right)\right].
\end{eqnarray}
We have already identified that 
$|\psi_{\rm ME}\rangle$ is a crucial AFS.
It seems reasonable to assume that AFSs
appearing before and after this state, having anomalously large 
fluctuations of $\hat M_x$,  
should also be crucial AFSs.
Figures \ref{yuragi15_1} - \ref{yuragi20_2} 
suggest that 
the number of such AFSs is roughly 
proportional to the total number of computational steps, $Q$.
As described in subsection \ref{ss-cs}, 
$Q=O(L_1^n)=O(L^n)$, where
$n=2$ in our simulation, whereas $n=3$ in an actual computation.
Hence, for large $L$, we can estimate roughly as
\begin{equation}
T \simeq
T_{\rm clean} \exp 
\left[
- \lambda^2 O(L^{n+2})
\right].
\label{scalingT}\end{equation}
Hence, 
$\epsilon$ defined by Eq.\ (\ref{epsilon}) is roughly given by
\begin{equation}
\epsilon \simeq
\exp 
\left[
- \lambda^2 O(L^{n+2})
\right].
\label{scalingepsilon}\end{equation}
To get a successful result within a time 
of ${\rm poly}(\log N) = {\rm poly}(L)$, where poly denotes a polynomial, 
the lowest allowable value $\epsilon_{\rm th}$ 
of $\epsilon$ should be
\begin{equation}
\epsilon_{\rm th} = 1/{\rm poly}(L).
\end{equation}
Therefore, a sufficient condition would be
\begin{equation}
\lambda^2 = O\left( \frac{\log(L)}{L^{n+2}} \right). 
\label{conditon2}\end{equation}
For any value of $n$, 
we stress that the required condition for $\lambda$
becomes $L$ ($\gg 1$) times severer due to the appearance of AFSs than 
the case where only NFSs appear during the computation, 
and conjecture (iii) has been confirmed.

\subsection{Possible meaning of the appearance of a noncrucial AFS}

It seems that macroscopic entanglement is necessary for 
the exponential speedup over classical computers in performing 
computation with huge inputs, because
otherwise the quantum computation would be able to be 
simulated efficiently by classical computers.
Since AFSs have macroscopic entanglement, 
they seem necessary for 
the exponential speedup.

We have shown that both 
$|\psi_{\rm ME}\rangle$ and
$|\psi_{\rm final} \rangle$ in Shor's factoring algorithm are AFSs.
In agreement with the general theory of SM \cite{SM02}, 
the decoherence rates of these states become anomalously great
as the number $L$ of qubits is increased.
However, we have also shown that 
the anomalously fast decoherence of 
$|\psi_{\rm final} \rangle$ is not crucial, i.e., 
does not reduce the success probability, 
whereas that of $|\psi_{\rm ME}\rangle$ is crucial.
This may indicate that the use of an AFS 
in the final stage would 
be dispensable, whereas
the use of an AFS in the modular exponentiation stage 
would be indispensable, to Shor's factoring algorithm; namely, 
it might be possible to modify 
the algorithm in such a way that it does not use an AFS in 
the final stage, while keeping the 
number $Q$ of computational steps as 
$Q={\rm poly}(\log N)= {\rm poly}(L)$.
On the other hand, 
it would be impossible to modify 
the algorithm 
in such a way that it does not use an AFS in 
the modular exponentiation stage, while keeping $Q = {\rm poly}(\log N)$.
This interesting point will be a subject of future studies.

\subsection{A possible method of fighting against anomalously fast 
decoherence of crucial AFSs}

If the use of an AFS in the 
modular exponentiation stage 
is indeed indispensable to Shor's algorithm,
we must find a way of fighting against the anomalously fast 
decoherence of such crucial AFSs.
A possible solution may be the use of ECC \cite{ec1,ec2,ec3,ec4}
and/or a DFS \cite{dfs1,dfs2,dfs3}, 
which will be 
mentioned in Sec. \ref{ss-ECandDFS}.
We here point out another possible solution.

We note that, according to formula (\ref{r-noise-AFS}),
an AFS becomes fragile only when the spectrum density of 
the noise behaves as $g(k_0) = O(L^{-1+\delta})$ with $\delta >0$. 
That is, when
$| \psi \rangle$ is an AFS for which 
$\langle \psi | \Delta \hat A_{k_0}^\dagger \Delta \hat A_{k_0} | \psi \rangle
= O(L^0)$, 
and when the spectrum density of the noise at $k=k_0$ 
behaves as $g(k_0) = O(L^{-1+\delta})$, 
then $| \psi \rangle$ becomes fragile if $\delta >0$
or non fragile if $\delta \leq 0$.
For example, if the physical dimension of the quantum computer is $1$ cm, 
and if the wavelengths of all noises are longer than $1$ cm, 
the spectral densities of the noises behave like
Eq.\ (\ref{longwlnoise}).
In such long-wavelength noises, AFSs with $k_0 = 0$ become fragile, 
whereas AFSs with $k_0 \gg 1 {\rm cm}^{-1}$ are nonfragile. 
Therefore, we would be able to avoid anomalously-fast decoherence  
in such long-wavelength noises if we can improve the algorithm 
in such a way that crucial AFSs are replaced with other 
AFSs with $k_0 \gg 1 {\rm cm}^{-1}$.

For example, 
we have shown that $|\psi_{\rm ME}\rangle$ is a crucial 
AFS that is fragile 
in a long-wavelength, bit-flip noise.
In order to realize quantum computers with large $L$, 
one should improve the algorithm 
in such a way that $|\psi_{\rm ME}\rangle$ is replaced with another 
AFSs with $k_0 \gg 1 {\rm cm}^{-1}$.
Since the construction of such an improved algorithm 
is beyond the scope of the present paper, 
we will study it elsewhere.

\subsection{Possibilities of other AFSs and other noises}
\label{ss-other}

As normalized additive operators, we have only examined the
``magnetizations'' $\hat M_\alpha$ and $\hat M_\alpha^{(1)}$ with 
$\alpha = x,y,z$.
This is sufficient for the purpose of the present paper, 
as explained in Sec. \ref{ss-choice}.
However, it would be interesting to find out all AFSs used in the 
algorithm.
To do this, we must study fluctuations of {\em all} 
normalized additive operators.
A convenient method of doing this has recently been
proposed by Sugita and Shimizu, and has been applied to 
quantum chaotic systems \cite{SS03}.
It would be very interesting to apply their method to 
quantum computers, and thereby make a complete list of 
anomalous states used in the quantum computation.

To examine the anomalously fast decoherence of
AFSs with anomalously large fluctuations of $\hat M_\alpha$, 
we have considered long-wavelength noises.
In real systems, it seems reasonable to assume that 
many different noises and/or components (wavelengths, frequencies, 
and directions) coexist. 
It would be necessary to examine effects of all existing 
noises to realize a quantum computer with a huge number $L$ of qubits,
because, as we have shown in this paper, for huge $L$ 
some noises can be crucial even if its strength is much weaker than 
other noises.

These points will be subjects of future studies.

\subsection{Uses of error correcting codes 
and/or a decoherence-free subspace}
\label{ss-ECandDFS}

We have studied Shor's factoring algorithm 
without ECC \cite{ec1,ec2,ec3,ec4} or
error avoiding using a DFS \cite{dfs1,dfs2,dfs3}; 
namely, we have studied ``bare'' characteristics
of quantum computers performing Shor's algorithm.

Regarding the use of a DFS, it will be effective
for fighting against 
the long-wavelength noises $f_\alpha$ that 
are considered in this paper.
However, as mentioned in the preceding section, 
many different noises and/or components 
would coexist in real systems. 
It is not clear whether a DFS can be constructed
efficiently in such a realistic case,
because the number of extra qubits that are necessary to construct 
a DFS is increased as the number of different 
noises and/or components is increased \cite{dfs1,dfs2}.
Also, 
at present, we do not have definite 
conclusions on the efficiency of 
ECC in such a realistic case.
We can, however, say that the improvement of 
the bare characteristics is crucial in practical applications,
because if the bare characteristics are bad, then 
ECC would become much complicated and large scale.
Such a computer system would be impractical.

Therefore, we think that all of ECC, a DFS, and
the improvement of the bare characteristics should be 
necessary to realize a quantum computer
that accept huge inputs.

\subsection{Various definitions of decoherence}

The decoherence and decoherence rate 
can be defined in many different ways. 
The definition of $\Gamma_m$ of the present paper, 
Eq.~(\ref{Gamma_m}), 
is one of many possible definitions.
Fortunately, however, we are interested in the case of small
$\lambda$ in this paper.
In such a case, we can obtain similar conclusions 
for many different definitions of the decoherence rate.
Indeed, we have shown in Sec. \ref{ss-fd} 
that the loss of fidelity and 
the increase of the $\alpha$ entropy agree with each other 
for small $\lambda$.

Note, however, that the situation could be different 
if one employed some definitions used in condensed-matter physics.
In condensed-matter physics, 
decoherence is often discussed 
with respect to a particular quantity, 
such as the line shape \cite{AW} and 
a nonequilibrium noise \cite{SU}.
In quantum information theory, on the other hand, 
the decoherence is usually 
defined with respect to all possible observable. 
The decoherence in the latter sense
can take place without any change of 
the (expectation) value of a particular quantity.
This point should be considered 
when using rich results of condensed-matter physics.


\section{summary and conclusions}

With the help of the general theory of stabilities of many-body quantum 
states by SM \cite{SM02}, 
we have studied properties of quantum states in quantum computers 
that perform Shor's factoring algorithm \cite{Shor_1}.
Since the efficiency of computation becomes relevant only when 
the size $N$ of the input is huge, we focus on asymptotic 
behaviors as the number $L$ [$=O(\log N)$] of qubits is increased. 

Following the general theory by SM, we have paid attention to 
fluctuations of normalized additive operators,
which are the sums of local operators, as  
defined by Eq.\ (\ref{A}).
If fluctuation $\langle \psi | (\Delta \hat A)^2 | \psi \rangle$
of every normalized additive operator $\hat A$ is $O(1/L)$ or less for
a pure state $| \psi \rangle$, we call it a normally fluctuating state
(NFS).
Any separable state is a NFS, whereas 
the inverse is not necessarily true.
On the other hand, 
if there is a normalized additive operator whose
fluctuation is $O(L^0)$ for a pure state $| \psi \rangle$, 
we call $| \psi \rangle$ an
anomalously fluctuating state (AFS).
Since AFSs have such anomalously large fluctuations, 
they are entangled macroscopically.
We have pointed out in Sec. \ref{ss-NFSAFS} that 
the use of 
fluctuations of normalized additive operators
as a measure of macroscopic entanglement 
seems natural from the viewpoints of many-body physics
and experiments.

By performing numerical simulations of Shor's factoring algorithm, 
we have found that AFSs appear during the computation, although 
the initial state is a NFS.
For example, the state $| \psi_{\rm ME} \rangle$ just after 
the modular exponentiation process
is an AFS, which has 
anomalously large fluctuation of $\hat M_x$.
Here, $\hat M_x$ is the $\alpha = x$ component of
$\hat M_\alpha$ that is defined by Eq.\ (\ref{Ma}), which 
corresponds to 
the magnetization of a spin system.
The final state $| \psi_{\rm final} \rangle$, 
which is obtained by a discrete Fourier transform, 
is also an AFS, which has 
anomalously large fluctuation of $\hat M_z$
(the $\alpha = z$ component of $\hat M_\alpha$).

According to the general theory by SM, 
the asymptotic behavior, as $L$ is 
increased, of the decoherence rate $\Gamma$ of 
a quantum state is directly related to 
fluctuations of normalized additive operators:
$\Gamma$ of a NFS never exceeds
$O(L)$ in any weak classical noises, whereas 
$\Gamma$ of an AFS can be either 
$O(L^2)$ or less, depending on the spectral densities of the noises.
An AFS with an anomalously great decoherence rate
$\Gamma=O(L^{1+\delta})$, where $\delta>0$, is said to be fragile.
We have shown that AFSs which we have found in Shor's algorithm
become fragile, $\Gamma=O(L^2)$, in 
long-wavelength noises with a $1/\mathit{f}$ spectrum, which simulate
noises in some real systems.
Therefore, for large $L$, the decoherence rate of a quantum computer 
performing Shor's factoring algorithm is 
determined by the decoherence rates of the fragile AFSs
that appear during the computation.

Since decoherence of particular states does not necessarily lead to 
false results of the quantum computation, we
examine whether the anomalously fast decoherence of the fragile AFSs
leads to anomalously large degradation of the result of computation.
To do this,  
we have investigated effects of the noises on 
the computational results for each state.
We have found that 
the anomalously fast decoherence, $\Gamma=O(L^2)$, of 
$|\psi_{\rm ME}\rangle$ leads to 
anomalously large reduction, approximately proportional to $L^2$,
of the success probability $T$, 
which is defined as the probability 
of getting the correct result of computation.
We have concluded that 
the anomalously fast decoherence of 
such crucial AFSs becomes a bottleneck of quantum computers performing 
Shor's factoring algorithm.

This does not however mean that all AFSs appearing in the algorithm 
are crucial.
For example, 
the anomalously fast decoherence, $\Gamma=O(L^2)$, of 
$|\psi_{\rm final}\rangle$
does not lead to large reduction of $T$.
In this sense, $|\psi_{\rm final}\rangle$ is a noncrucial AFS
in Shor's factoring algorithm.

It seems that macroscopic entanglement is necessary for 
the exponential speedup over classical computers in performing 
computation with huge inputs, because
otherwise the quantum computation would be able to be 
simulated efficiently by classical computers.
Since AFSs have macroscopic entanglement, 
they seem necessary for 
the exponential speedup.
This does not however mean 
that all AFSs in an algorithm written naively 
are necessary.
For Shor's factoring algorithm, 
our results suggest that a crucial AFS 
$|\psi_{\rm ME}\rangle$ is necessary whereas 
a noncrucial AFS $|\psi_{\rm final}\rangle$ 
may be able to be replaced with a NFS.
In order to realize quantum computers with large $L$, 
one should improve the algorithm 
in such a way that necessary but fragile AFSs are replaced with other 
AFSs that are nonfragile in real noises.
To do this, formula (\ref{r-noise-AFS}) will be useful, 
from which one can estimate 
$\Gamma$ of an AFS.
Since error correcting codes and decoherence-free subspaces
are not almighty, 
we think that one must also utilize 
such optimization to realize a quantum computer that 
accepts huge inputs.

\begin{acknowledgments}
We thank T.\ Miyadera for fruitful discussions and comments.
This work was partly supported by 
Grant-in-Aid for Scientific Research.
\end{acknowledgments}


\begin{thebibliography}{99}

\bibitem{Deutsch_1}
         D.\ Deutsch,
         Proc.\ R.\ Soc.\ London,\ Ser.\ A\ {\bf 400},\ 97 (1985).

\bibitem{Shor_1}
        P.\ W.\ Shor,
	Proceedings of the 35th Annual Symposium on 
the Foundations of Computer Science, edited by S.\ Goldwasser 
(IEEE Computer Society, Los Alamitos, CA, 1994) p. 124.
       
        

\bibitem{Ekert_Jozsa}
         A.\ Ekert and R.\ Jozsa,
         Rev.\ Mod.\ Phys. {\bf 68}, 733 (1996).

\bibitem{Nielsen_Chaung}
        M.\ A.\ Nielsen and I.\ L.\ Chuang,
        {\it Quantum Computation and Quantum Information}
        (Cambridge University Press, Cambridge, 2000).
        

\bibitem{SM02}
A.\ Shimizu and T.\ Miyadera, 
Phys.\ Rev.\ Lett.\ {\bf 89}, 270403 (2002). 

\bibitem{haag}
R.\ Haag, {\it Local Quantum Physics} (Springer, Berlin, 1992).

\bibitem{unruh}
W.\ G.\ Unruh, Phys.\ Rev.\ A {\bf 51}, 992 (1995).

\bibitem{Palma}
         G.\ M.\ Palma,\ K.-A.\ Suominen,\ and A.\ K.\ Ekert,
         Proc.\ R.\ Soc.\ London,\ Ser.\ A\ {\bf 452},\ 567 (1996).

\bibitem{Miquel_Paz_Perazzo}{%
        C.\ Miquel,\ J.\ P.\ Paz and R.\ Perazzo,
        Phys.\ Rev.\ A\ {\bf 54},\ 2605 (1996).
        }

\bibitem{HL}
P.\ Horsh and W.\ von der Linden, Z.\ Phys.\ B: Condens. Matter {\bf 72}, 181 (1988).

\bibitem{KT}
T. Koma and H. Tasaki, 
J.\ Stat.\ Phys.\ {\bf 76}, 745 (1994).

\bibitem{pre01}
A.\ Shimizu\ and\ T.\ Miyadera, 
Phys.\ Rev.\ E {\bf 64}, 056121 (2001).

\bibitem{SS03}
A.\ Sugita and A.\ Shimizu, 
e-print: quant-ph/0309217.

\bibitem{disadv}
On the other hand, a disadvantage of the present measure 
of entanglement is that it can only be applied to pure states.

\bibitem{NGmodes}
For example, a ground state in which a continuous symmetry is 
broken often takes such a value of $p$ because of the Nambu-Goldstone modes.

\bibitem{k}
Since the noise amplitude outside the qubit system is irrelevant, 
the spatial Fourier transform is taken over 
$\ell = 1, 2, \cdots, L$ to define $g(k,\omega)$, 
hence $k$ takes discrete values with separation $2 \pi/L$.


\bibitem{miyake}
A. Miyake and M. Wadati,
Phys. Rev. A {\bf 64}, 042317 (2001).

\bibitem{Jozsa_Linden}
        R.\ Jozsa and N.\ Linden,
        e-print quant-ph/0201143.



\bibitem{ec1}
P. W. Shor, Phys. Rev. A {\bf 52}, 2493 (1995).

\bibitem{ec2}
A. Ekert and C. Macchiavello, Phys. Rev. Lett. {\bf 77}, 2585 (1996).

\bibitem{ec3}
D. Gottesman, Phys. Rev. A {\bf 54}, 1862 (1996).

\bibitem{ec4}
J.\ Preskill, Phys. Today, {\bf 52}(6), 24 (1999).

\bibitem{dfs1}
P. Zanardi and M. Rasetti, Phys. Rev. Lett. {\bf 79}, 3306 (1997).

\bibitem{dfs2}
D. A. Lidar, I. L. Chuang and K. B. Whaley,
Phys. Rev. Lett. {\bf 81}, 2594 (1998).

\bibitem{dfs3}
D. Kielpinski {\it et al}., 
Science {\bf 291}, 1013 (2001).

\bibitem{x}
One can check whether ${\sf x}$ is coprime to $N$
easily by a classical computer.

\bibitem{1fa}
M. Covington, M. W. Keller, R. L. Kautz, and J. M. Martinis, 
Phys.\ Rev.\ Lett.\ {\bf 84}, 5192 (2000). 

\bibitem{1fb}
R. L. Kautz, M. W. Keller, and J. M. Martinis,
Phys.\ Rev.\ B {\bf 62}, 15888 (2000). 

\bibitem{nakamura}
Y.\ Nakamura, Yu. A. Pashkin, T. Yamamoto, and J. S. Tsai,
Phys.\ Rev.\ Lett.\ {\bf 88}, 047901 (2002). 

\bibitem{1/f}
$1/{\it f}$ noises do not strictly satisfy the assumptions
of Ref.\ \cite{SM02}. 
This does not cause any difficulty in the present paper, 
as discussed at the end of subsection \ref{ss-fd}.

\bibitem{nonlinear}
Although the relation between $T_{\rm clean}-T_m$ and $1-F_m$  
becomes nonlinear for larger $\lambda$ and/or larger $L$, 
we confirmed that the relation is still monotonic.



\bibitem{AW}
P. W. Anderson and P. R. Weiss, 
Rev. Mod. Phys. {\bf 25}, 269 (1953).

\bibitem{SU}
A. Shimizu and M. Ueda, 
Phys. Rev. Lett. {\bf 69}, 1403 (1992): {\bf 86},
 3694(E) (2001).

\bibitem{S_talk}
A.~Shimizu, talk presented at The 4th Symposium on Quantum Effects 
and Related Physical Phenomena (December 20-21, 2000, Tokyo, Japan, unpublished); A.~Shimizu and T.~Miyadera, in {\it Proceedings of 
the 56th annual meeting of the Physical Society of Japan}
(Physical Society of Japan, 2001), paper no. 28pYN-6.
\end{thebibliography}
\end{document}